\documentclass[useAMS,usenatbib]{mn2e}
\usepackage[latin2]{inputenc} 
\usepackage[pdftex]{graphicx} 
\usepackage{hyperref} 
\usepackage{amsmath} 
\usepackage{amssymb} 
\usepackage{ulem} 
\usepackage{pdflscape}
\usepackage{subfig}
\usepackage{lipsum}
\usepackage{color,soul} 
\usepackage{multicol}
\usepackage{graphicx}
\usepackage[export]{adjustbox}
\usepackage{tabularx} 

\usepackage{layouts}
\graphicspath{{./figs/}}
\newcommand{\HI}{H\,{\sc i} }

\newcommand{\HInospace}{H\,{\sc i}}

\newcommand{\MHInospace}{M$_{\rm HI}$}
\newcommand{\Mstar}{M$_{\star}$ }
\newcommand{\Mstarnospace}{M$_{\star}$}

\newcommand{\squiggle}{$\mathtt{\sim}$}
\newcommand{\mustar}{$\mu_{\star}$ }
\newcommand{\Msol}{M$_{\odot}$}

\makeatletter
\newsavebox\myboxA
\newsavebox\myboxB
\newlength\mylenA

\newcommand*\xoverline[2][0.75]{%
    \sbox{\myboxA}{$\m@th#2$}%
    \setbox\myboxB\null
    \ht\myboxB=\ht\myboxA%
    \dp\myboxB=\dp\myboxA%
    \wd\myboxB=#1\wd\myboxA
    \sbox\myboxB{$\m@th\overline{\copy\myboxB}$}
    \setlength\mylenA{\the\wd\myboxA}
    \addtolength\mylenA{-\the\wd\myboxB}%
    \ifdim\wd\myboxB<\wd\myboxA%
       \rlap{\hskip 0.5\mylenA\usebox\myboxB}{\usebox\myboxA}%
    \else
        \hskip -0.5\mylenA\rlap{\usebox\myboxA}{\hskip 0.5\mylenA\usebox\myboxB}%
    \fi}
\makeatother

\newcolumntype{b}{X}
\newcolumntype{s}{>{\hsize=.5\hsize}X}

\title[Gas stripping in satellite galaxies]
{Cold gas stripping in satellite galaxies: from pairs to clusters}
\author[Toby Brown et al.]
{Toby Brown$^{1,2}$\thanks{thbrown@swin.edu.au},
Barbara Catinella$^2$,
Luca Cortese$^2$, 
Claudia del P. Lagos$^{2,3}$,
\newauthor
Romeel Dav\'{e}$^{4,5,6,7}$, 
Virginia Kilborn$^1$, 
Martha P. Haynes$^8$,
Riccardo Giovanelli$^8$
\newauthor  and Mika Rafieferantsoa$^{4,5}$
\\
$^1$Centre for Astrophysics and Supercomputing, Swinburne University of Technology, Hawthorn, VIC 3122, Australia\\
$^2$International Centre for Radio Astronomy Research, University of Western
Australia, 35 Stirling Highway, Crawley, WA 6009, Australia\\
$^3$ARC Centre of Excellence for All-sky Astrophysics (CAASTRO), 44 Rosehill Street, Redfern, NSW 2016, Australia\\
$^4$Physics Department, University of Western Cape, Bellville, Cape Town 7535, South Africa\\
$^5$South African Astronomical Observatory, PO Box 9, Observatory, Cape Town 7935, South Africa\\
$^6$African Institute of Mathematical Sciences, Muizenberg, Cape Town 7945, South Africa\\
$^7$Astronomy Department and CCAPP, Ohio State University, Columbus, OH 43210, USA\\
$^8$Cornell Center for Astrophysics and Planetary Science, Space Sciences Building, Cornell University, Ithaca, NY 14853, USA}

\begin{document}
\date{Accepted 20XX. Received 20XX; in original form 20XX}

\pagerange{\pageref{firstpage}--\pageref{lastpage}} \pubyear{2002}

\maketitle

\label{firstpage}

\begin{abstract}
In this paper we investigate environment driven gas depletion in satellite galaxies, taking full advantage of the atomic hydrogen (\HInospace) spectral stacking technique to quantify the gas content for the entire gas-poor to -rich regime. We do so using a multi-wavelength sample of 10,600 satellite galaxies, selected according to stellar mass (log \Mstarnospace/\Msol \space $\geq$ 9) and redshift ($0.02 \leq {\it z} \leq 0.05 $) from the Sloan Digital Sky Survey, with \HI data from the Arecibo Legacy Fast ALFA (ALFALFA) survey. Using key \HInospace-to-stellar mass scaling relations, we present evidence that the gas content of satellite galaxies is, to a significant extent, dependent on the environment in which a galaxy resides. For the first time, we demonstrate that systematic environmental suppression of gas content at both fixed stellar mass and fixed specific star formation rate (sSFR) in satellite galaxies begins in halo masses typical of the group regime (log M$_{\rm h}$/\Msol \space $<$ 13.5), well before galaxies reach the cluster environment. We also show that environment driven gas depletion is more closely associated to halo mass than local density. Our results are then compared with state-of-the-art semi-analytic models and hydrodynamical simulations and discussed within this framework, showing that more work is needed if models are to reproduce the observations. We conclude that the observed decrease of gas content in the group and cluster environments cannot be reproduced by starvation of the gas supply alone and invoke fast acting processes such as ram-pressure stripping of cold gas to explain this.

\end{abstract}


\section{Introduction}
\label{sec:Introduction}
Galaxies are vast collections of gravitationally bound stars and gas, situated inside haloes of dark matter (DM). It is known that one of the key mechanisms by which these gigantic systems perpetuate themselves is the conversion of their neutral atomic hydrogen (\HInospace) into stars via the transitional molecular phase (H$_{2}$). In light of this, if we are to paint a complete picture of galaxy formation and evolution, it is important that we seek to fully understand the complex links between cold gas reservoirs, their parent galaxy and the DM haloes that host them.

The majority of observational \HI studies have historically relied upon relatively small sample sizes in comparison to works at other wavelengths. In recent times, large area, blind surveys such as the Arecibo Legacy Fast ALFA \citep[ALFALFA;][]{Giovanelli2005a} survey have begun to address this. In its latest data release \citep[$\alpha.70$\footnote{\url{http://egg.astro.cornell.edu/alfalfa/data/}};][]{Haynes2011}, ALFALFA provided global \HI measurements for more than twenty-three thousand galaxies in the nearby Universe. In addition to blind-surveys, the GALEX Arecibo SDSS Survey \citep[GASS;][]{Catinella2010} provides the largest complement of targeted observations for a stellar mass selected sample, probing almost one thousand galaxies (log \Mstarnospace/\Msol \space $\geq$ 10) across the entire gas fraction spectrum.

The \HI spectral stacking technique allows studies to extend their scope beyond the sensitivity limits of current surveys. By optically-selecting and co-adding \HI spectra, \citet{Fabello2011} and \citet{Brown2015} have investigated the main gas scaling relations, probing lower \HI content and increased sample sizes beyond what is currently possible using only detections. There are other applications of the \HI stacking technique, examples include studies by \citet{Gereb2015} who use stacking to probe \HI content out to z \squiggle0.1 and \citet{Lah2009} who investigate variance in the cosmic density of \HI ($\Omega_{\rm HI}$) with redshift.

Using both detections and stacking, there have been many investigations into the relationships between galaxy properties and \HI content. Studies have established strong links between gas content and star formation density \citep{Schmidt1959,Kennicutt1998}, stellar mass \citep{Gavazzi1996} and morphology \citep{McGaugh1997}. However, galaxies are not isolated systems, for example \citet{Dressler1980} showed that the morphological fraction changes dramatically with the density of galaxies, thus understanding the influence of environment upon their evolution has become increasingly important. Studies of cold gas (\squiggle10$^2$ K) show that reservoirs are adversely affected in the highest density environments such as galaxy clusters \citep{Giovanelli1985,Chung2009,Cortese2011,Serra2012}, and that gas processing begins to occur within the group environment \citep{Kilborn2009,Rasmussen2012,Catinella2013,Hess2013}. Throughout the literature, depletion of \HI content due to environment is attributed to several different processes: the interaction between the interstellar-medium (ISM) and intergalactic-medium (IGM) known as ram-pressure stripping \citep{Gunn1972,Hester2006}; heating and stripping of hot gas in the DM halo preventing replenishment \citep[strangulation;][]{Larson1980}; high \citep[harassment;][]{Moore1998} and low \citep[tidal stripping;][]{Moore1999} velocity gravitational interactions with neighbours.

At other wavelengths, work has built upon \citet{Dressler1980}, showing the strong dependencies of star formation \citep{Balogh1999,Gomez2003,Cooper2008} and morphology \citep{Whitmore1991,Poggianti2008,Wilman2012} upon environment. Disentangling the primary, secondary and, in some cases, tertiary connections between internal galaxy properties, environment and gas content is a topic of much current interest.

In pursuit of this, recent works have begun to separate and classify galaxies based upon their status as a central (most-massive and/or luminous) or satellite galaxy within the halo. The physical motivation is that satellites infalling into the halo have undergone a distinct evolutionary history from that of their central, as well as being the bulk of the group and cluster populations \citep[e.g. ][]{vandenBosch2008,Kimm2009,Wetzel2012,Peng2012,Woo2013}. Following this, \citet{vandenBosch2008} and \citet{Wetzel2012} argue that the environmental relationships and build-up of the red sequence are primarily driven by the quenching of satellites rather than their central and it is likely that the transformations are caused by removal or consumption of the cold gas content.

Despite such studies, there remains a paucity of work investigating the gas content of satellite galaxies from a statistical perspective. The extent to which \HI loss may be attributed to environment, what processes are at work and in which regimes are questions that have not yet been answered \citep{Yoon2015}. In this work we use the largest representative sample of \HI to date, coupled with the spectral stacking technique, to address these questions. We look to provide the very first large-scale, statistical census of cold gas and environment for satellite galaxies in the local Universe.

Section \ref{sec:Sample} contains an overview of the sample selection, environmental measures and stacking technique used in this paper. Sections \ref{sec:ScalingRelations} studies the main \HInospace-to-stellar mass ratio scaling relations as a function of halo mass. In Section \ref{sec:LD_scalingRelations} we characterise environment using nearest neighbour and fixed aperture densities, investigating their impact on gas fraction. We compare our results with theoretical predictions in Section \ref{sec:models}. Lastly, Section \ref{sec:Discussion} discusses our conclusions and considers the physical mechanisms at play.

Throughout this paper the distance dependent quantities computed using observations assume a $\Lambda$CDM cosmology with $\Omega = 0.3$, $\Lambda = 0.7$, and a Hubble constant $\text{H}_{0} = 70\; \text{km s}^{-1} \;\text{Mpc}^{-1}$. Where theoretical models and simulations are used for comparison the relevant cosmological parameters are listed in the description.

\section[]{The Sample}
\label{sec:Sample}
\subsection{Selection}
The findings presented in this paper are based upon a subset of the {\it parent sample}, hereafter Sample I, of 30,695 galaxies defined in Section 2 of \citet[][hereafter Paper I]{Brown2015}. Briefly, Sample I is a volume-limited ($0.02 \leq {\it z} \leq 0.05 $), stellar mass selected (9 $\leq$ log M$_{\star}$/M$_\odot \leq$ 11.5) sample drawn from the overlap between the Sloan Digital Sky Survey Data Release 7 \citep[SDSS DR7;][]{Abazajian2009} footprint and the sky area of the ALFALFA survey processed to date ($112.5\textdegree \leq \alpha \leq 247.5\textdegree $; $0\textdegree \leq \delta \leq 18\textdegree $ \& $24\textdegree \leq \delta \leq 30\textdegree $).

Optical data in this work are based upon SDSS {\it ugriz} model magnitudes - the optimal fit of a de Vaucouleurs or exponential profile to the flux in each band - corrected for Galactic extinction \citep[see ][]{Schlegel1998}. \HI data are in the form of 21-cm line emission spectra extracted from ALFALFA data cubes at the position of our target galaxies using SDSS coordinates. In this way \HI spectra are provided for the full sample of galaxies regardless of their detection status. For a full breakdown of the radio and optical data the reader should refer to Sections 2.2 and 2.3 of Paper I.

Stellar masses, derived via photometric fitting, and SFR estimates are taken from the value-added MPA-JHU SDSS DR7
\footnote{\url{http://wwwmpa.mpa-garching.mpg.de/SDSS/DR7/}
using improved stellar masses from \url{http://home.strw.leidenuniv.nl/~jarle/SDSS/}} 
catalogue. For both these quantities, the median values of the probability density function are used. Star formation rates are calculated following the methodology of \citet{Brinchmann2004} who use H$\alpha$ emission line modelling where available (S/N $>$ 3) for star forming galaxies. Where no or low S/N emission lines are present SFRs are computed using the empirical relationship between specific SFR (sSFR = SFR/M$_{\star}$) and the strength of the break at 4000\space\AA \space(D4000). Aperture corrections are applied using stochastic fits to the photometry \cite[see ][]{Salim2007} and global SFRs are recovered.

\subsection{Measures of Galaxy Environment}
There are many different metrics by which one may define the environment of a galaxy, most of which have been examined extensively in the literature \citep[see][for a thorough comparison]{Muldrew2012}. The majority of methods can be placed into the categories of friends-of-friends (FoF), {\it Nth} nearest neighbour and fixed aperture techniques. In general, approaches that are based upon FoF estimate properties such as mass and velocity dispersion, attributes closely related to gravitational potential. On the other hand, nearest neighbour or fixed aperture estimators provide the straight up number density of objects, a property that is indicative of the probability of interaction. We employ one metric from each of these categories in order to best determine the extent and influence of processes at work as well as the subsequent scale dependency of environment driven gas suppression.

We divide the sample cleanly into central (most massive galaxy in group), isolated (only galaxy in group) and satellite (less massive than central in groups of two or more members) galaxies based upon the \citet{Yang2007}\footnote{\url{http://gax.shao.ac.cn/data/Group.html}} group catalogue, hereafter Y07. For the analysis in this paper, we focus {\it only} on the satellites and restrict the parent sample to objects for which the full complement of environment data outlined below is available, 10,567 in total.

Below we describe the three environment metrics used in this paper.

\subsubsection{Friends-of-Friends and Halo Based Group Catalogue}
\label{sec:FoF}
The principle of FoF algorithms is that galaxies are associated with one another based upon their spacial proximity, defining a group using all objects at a proximity less than a given linking length. The advantage of FoF is that, once the group is defined, secondary derived properties such as mass and velocity dispersion may be assigned.

In this paper, the DM halo masses for Sample I galaxies are provided by Y07. More specifically, we use the `modelB' version which takes model magnitudes along with redshift measurements taken from SDSS, however, when necessary, it also uses redshifts from additional surveys \citep[e.g. 2df;][]{Colless2001}. The authors apply the halo-based, FoF group finder algorithm developed by \citet{Yang2005} to SDSS DR7. The basic procedure assigns centres to potential groups and assumes an initial mass-to-light ratio, allocating a provisional mass to each group using their characteristic luminosity. They then use this provisional mass to estimate the size and velocity dispersion of the host DM halo, using these properties to determine a density contrast for each halo and assign galaxies to their most likely group. The process is repeated until the group membership stabilises and the resulting catalogue is largely independent of the initial mass-to-light assumption.

Once the galaxy group association is confirmed, final halo masses are estimated by abundance matching the characteristic luminosity or stellar mass rank order of individual groups with the halo mass function of \citet{Warren2006}. No halo masses are assigned to the smallest groups (log M$_{\rm h}$/M$_\odot$ $<$ 11.6) in Y07. For our work we use halo masses based upon the stellar mass ranking and have estimated values below log M$_{\rm h}$/M$_\odot =$ 11.6 using an extrapolation of the mean relation between the stellar mass of the central galaxy and the parent halo mass provided by \citet[][Eqn. 7]{Yang2008}. Our lowest mass bin is log M$_{\rm h} < 12$, in this way direct comparison between manually assigned DM masses is avoided.

In some cases, the process of abundance matching may yield halo mass values that deviate significantly from the ``true'' halo mass \citep[see][]{Duarte2015}. We estimate this bias by applying the abundance matching method of \citet{Yang2007} to the GALFORM semi-analytic model \citep{Gonzalez2014} described in Section \ref{sec:models}. We find that estimated halo masses are, on average, 0.2 dex lower than than true masses with a standard deviation of 1 dex, 0.7 dex and 0.3 dex at true halo masses of log M$_{\rm h}$/M$_\odot =$ 12, 13.5 and 14.5 respectively. This spread is due to scatter in the predicted stellar mass-halo mass relation of GALFORM \citep{Guo2016,Mitchell2016}, meaning that stellar mass is not necessarily a clean predictor of halo mass. That said, the correlation is also significantly dependent on the implementation of feedback in the models and, as a consequence, other models produce a tighter stellar mass-halo mass relation (see \citealt{Guo2016} for a comparison of different models). The scatter introduced by abundance matching is smaller than any halo mass bin used in this paper.

At low redshift, massive groups and clusters exhibit an IGM hot enough to emit X-ray light via thermal bremsstrahlung radiation. This means that, where X-ray observations exist, the virialized nature of these systems can be independently confirmed and objective comparison can be made between their optical and X-ray derived properties. To this effect, \citet{Wang2011} identify 201 clusters in the Y07 catalogue between $0.01 \leq {\it z} \leq 0.2$ with counterparts from the combined ROSAT all sky survey X-ray cluster catalogues \citep{Ebeling1998,Ebeling2000,Bohringer2000,Bohringer2004}. Encouragingly, they find reasonable agreement between halo masses derived via abundance matching in Y07 and those calculated using X-ray cluster luminosity scaling relations. \citet{Wang2011}, and their subsequent paper \citet{Wang2014}, show that the X-ray luminosity is correlated with the total stellar mass of the cluster and the stellar mass of the central galaxy. The agreement between our chosen group catalogue and X-ray cluster observations, in addition to our testing of the abundance matching method, means that we can assume the halo masses of Y07 are a reliable estimate of the virial mass of large groups and clusters in our sample. We caution that, for small groups where multiplicity is low (${\rm N}_{\rm gal}\lesssim3$), the assumption of dynamic equilibrium may be incorrect. Although \citet{Yang2007} attempt to estimate halo masses for these objects in a meaningful manner via semi-empirical comparisons with mock catalogues, this remains an intrinsic problem with all group finding analyses and can affect the accuracy of the assigned halo masses in this regime.

\begin{figure}
    \includegraphics[width=\linewidth]{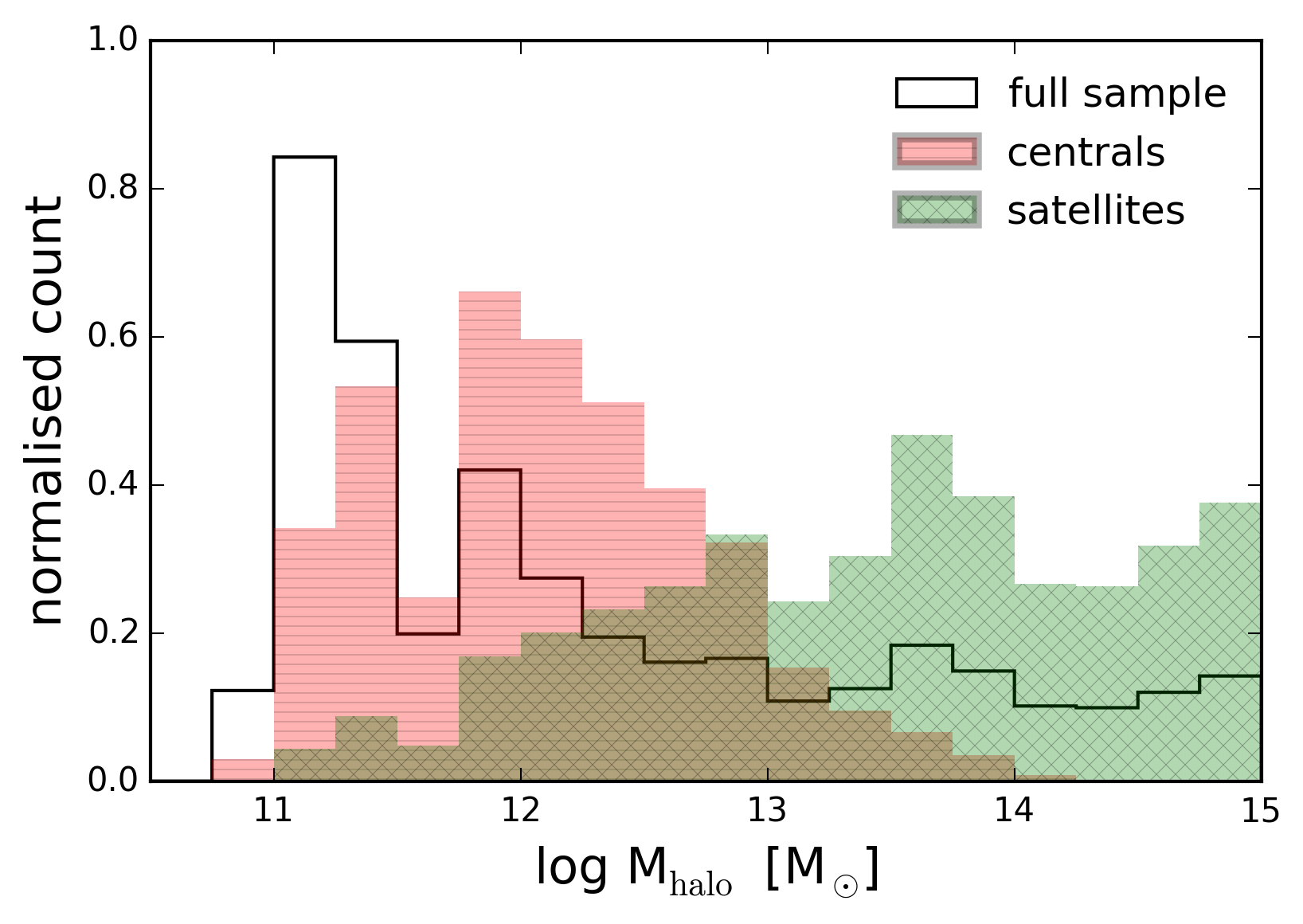}\par
  \caption{Group halo mass distribution from Y07 for galaxies within Sample I. The black solid line denotes the full sample of \squiggle27,700 galaxies while the red horizontal shaded histogram and green diagonally hatched histogram show the halo mass distribution of central (\squiggle2,800) and satellite (\squiggle10,600) galaxies respectively. Isolated galaxies (\squiggle14,300) are not shown.}
   \label{fig:mh_hist}
\end{figure}

Figure \ref{fig:mh_hist} shows the distribution of halo masses for the galaxies that are present in both Sample I and Y07. The sample of 27,667 galaxies is denoted by the solid black line while the 2,792 central and 10,567 satellite galaxies are displayed by the red and green shaded histograms respectively. 

\subsubsection{{\it Nth} Nearest Neighbour}
Simply put, the motivation for using the Nth nearest neighbour method is that galaxies in close proximity to their neighbours, by definition, reside in dense regions of the Universe. The closer a galaxy's neighbours the denser its environment.

We adopt a two dimensional nearest neighbour routine that applies a recessional velocity cut of $\pm$ 1000 km s$^{-1}$ around the target galaxy, calculates the distance of its {\it Nth} neighbour and defines the density of the subsequent volume as
\begin{flalign}
&\Sigma_{\rm N} = \dfrac{\rm N}{\pi r_{\rm N}^{2}}&
\label{eq:mu_star}
\end{flalign}
where N is the number of neighbours and {\it r}$_{\rm N}$ is the projected distance from the target galaxy to the Nth neighbour in kpc. 

We apply this to DR7, using a larger volume that encompasses Sample I and all galaxies above log \Mstarnospace/\Msol \space = 9. This ensures completeness and removes edge effects on Sample I galaxies near the volume boundaries. Within the literature, there is no clear consensus as to the optimal number of neighbours and, as discussed in \citet{Muldrew2012}, the decision depends on the scales one wishes to probe. We have investigated the differences in using 3rd, 5th, 7th and 10th nearest neighbour methods and find that while increasing N probes larger scales there is a strong correlation between all the neighbour-based methods. We find that the 7th nearest neighbour density \citep[e.g.][]{vanderWel2008,Muldrew2012} is a suitable match for the length scales present in Sample I, however the choice of N=7 is rather arbitrary and does not significantly affect our results.

\subsubsection{Fixed Aperture}
The fixed aperture technique is similar in concept to nearest neighbour, however, instead of defining a volume based upon distance to the Nth neighbour, one determines the number density of galaxies within a cylindrical volume of a given projected radius and velocity cut. Velocity cuts are intended to match the largest possible contribution of peculiar velocities to the sample, reducing interlopers that may be incorrectly placed into or out of an aperture. We compute the fixed aperture density on the larger volume used in the nearest neighbour method above. In order to match the large scales and peculiar velocities within sample I, we set the radius of our fixed aperture at 1 Mpc ($\pm$ 1000 km s$^{-1}$) centred on the galaxy \citep[e.g.][]{Grutzbauch2011,Muldrew2012}. We have also compared the effects of using apertures of  1 Mpc ($\pm$ 500 km s$^{-1}$), 2 Mpc ($\pm$ 1000 km s$^{-1}$) and 2 Mpc ($\pm$ 500 km s$^{-1}$). Our results do not depend significantly on the aperture choice.

The final sample used in this work contains 10,567 satellite galaxies (38 per cent of Sample I) for which there are observed atomic hydrogen and optical data along with subsequently derived stellar masses and star formation rates. Each galaxy has an assigned halo mass, as well as calculated nearest neighbour and fixed aperture densities. It is a sample built to be representative of the local Universe and therefore an ideal resource for studying environment driven evolution in the gas content of satellites from groups to clusters.

\subsection{Stacking and Errors}
\label{sec:Errors}
The focus of this work is to probe the relationship between \HI and environment in satellite galaxies. As discussed in the introduction, sensitivity limitations of current \HI surveys make it infeasible to obtain detections for very large, representative samples such as ours. This effect is particularly pronounced in studies of environment because of the \HI deficiency of galaxies found in the large group and cluster regimes \citep{Giovanelli1985,Cortese2011}. Using \HI stacking, we are able to quantify the gas content for the entire gas-poor to -rich regime and obtain the average atomic hydrogen content for each selection of co-added galaxies, regardless of whether the objects are formal detections in emission. We use an updated version of the stacking method developed by \citet{Fabello2011}, described fully in Section 3 of Paper I. Our sample consists of 1627 satellite galaxies detected in \HI (15 per cent) and 8940 non-detections (85 per cent).

One caveat of this technique is that, because of the use of spectral non-detections, the distribution of individual \HI masses from each stacked ensemble is not recoverable. Therefore the errors presented in this work are calculated using the statistical delete-a-group jackknife routine which iteratively discards a random 20 per cent of the stack selection, recomputing the average gas fraction. We do not allow any spectra to be used more than once, hence, this provides five independent stacks on which to compute the error. The jackknifed uncertainty is essentially the standard error on the mean gas fraction calculated by stacking and depends most strongly on the number of galaxies in each stack. This is discussed more thoroughly in Section 3.1 of Paper I.

Lastly, the rate of spectroscopic confusion with a sample is an important concern with all single-dish radio observations and in particular when stacking. Across the redshift range of Sample I, the rate of confusion within the ALFALFA dataset is no more than ten per cent \citep{Jones2015}. This is an acceptably low rate and unlikely to heavily bias the stacked average results. Further to this, although the number of sources blended within the Arecibo beam (3.5 arcmin) clearly increases in crowded environments, the impact of confusion is to increase stacked \HI mass and, thus, will not contribute to any observed trends of decreasing gas content with environment.

\begin{table}
\centering
\caption{Environments and their equivalent halo mass interval used throughout Section \ref{sec:ScalingRelations}. The upper and lower bin bounds are given in the first column. $\xoverline{\rm N}_{\rm gal}$is the mean number of group members found in each environment, while $\widetilde{\rm N}_{\rm gal}$ is the median value.}
\label{tab:env_class}
\resizebox{\linewidth}{!}{%
\begin{tabular}{cclll}
 \hline
x = log M$_{\rm h}$/M$_\odot$ & Environment & $\xoverline{\rm N}_{\rm gal}$ & $\widetilde{\rm N}_{\rm gal}$
\rule{0pt}{2.6ex}\rule[-0.9ex]{0pt}{0pt}\\ 
\hline

\hspace{2.4em}x $<$ 12  & Pairs/small groups  & 2 & 2\\
12 $\leq$ x $<$ 13          & Medium groups   & 5 & 4\\
13 $\leq$ x $<$ 14    & Large groups  & 26 & 21\\
\hspace{2.4em}x $\geq$ 14   & Clusters    & 242 & 169\\
\hline
\end{tabular}
}
\end{table}

\section{The Influence of Halo Mass upon Gas Fraction}
\label{sec:ScalingRelations}

\begin{figure*}
\begin{center}
\resizebox{\textwidth}{!}{%
\includegraphics[trim={0 0 1.2cm 0},clip, valign=t]{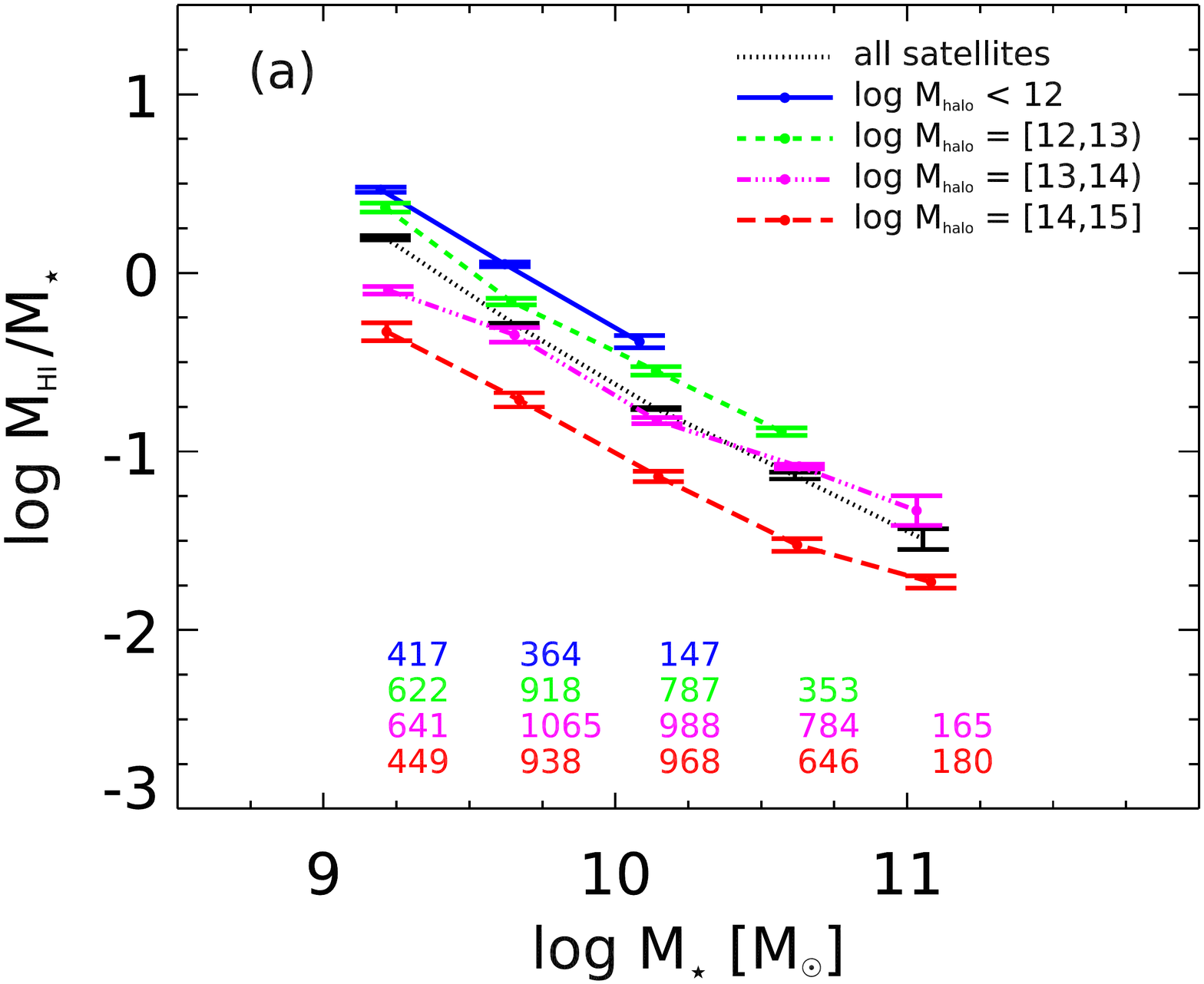}%
\hspace{-1cm}
\includegraphics[trim={3.8cm 0 1.2cm 0},clip, valign=t]{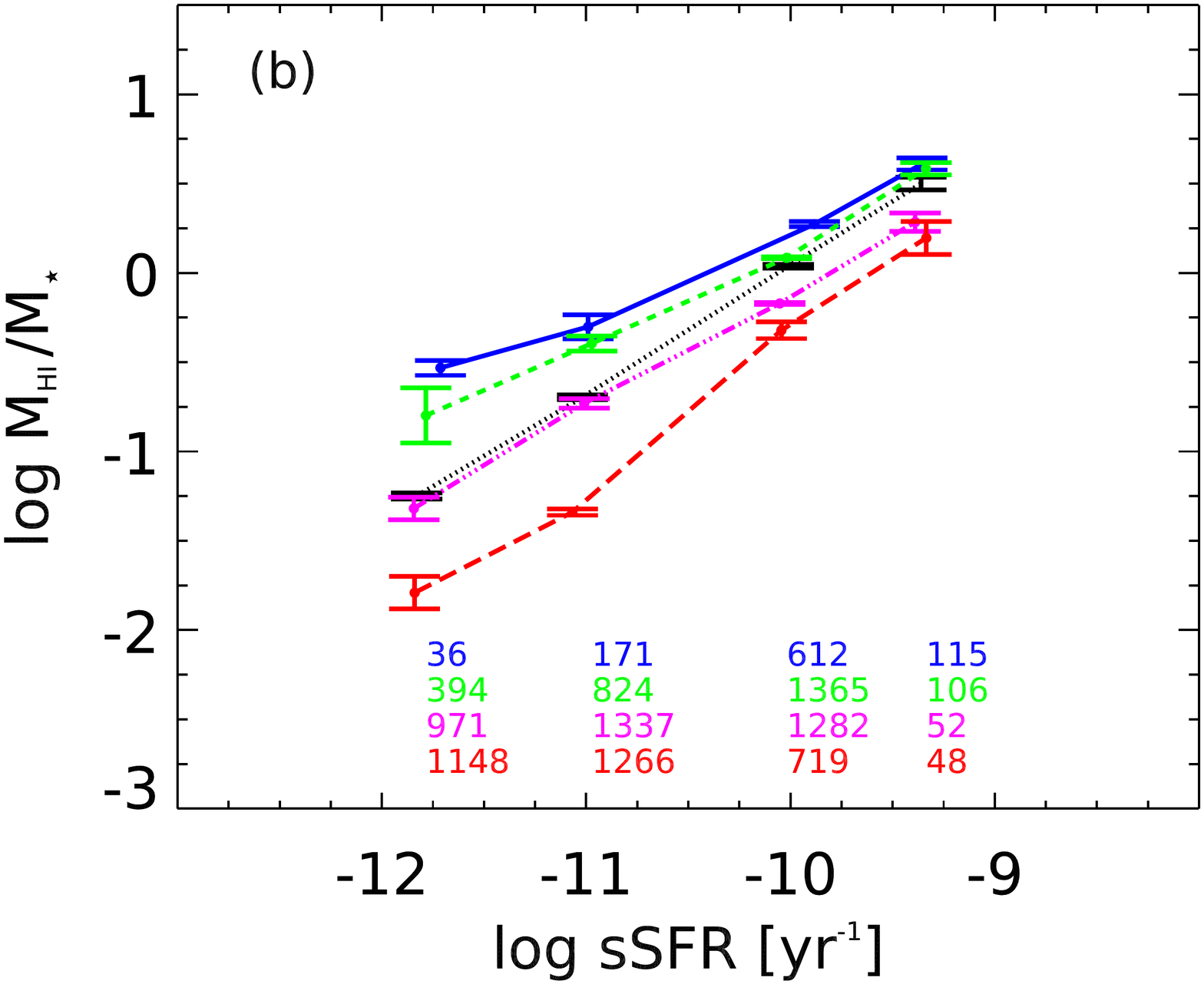}%
\hspace{-1cm}
\includegraphics[trim={3.8cm 0 1.2cm 0},clip, valign=t]{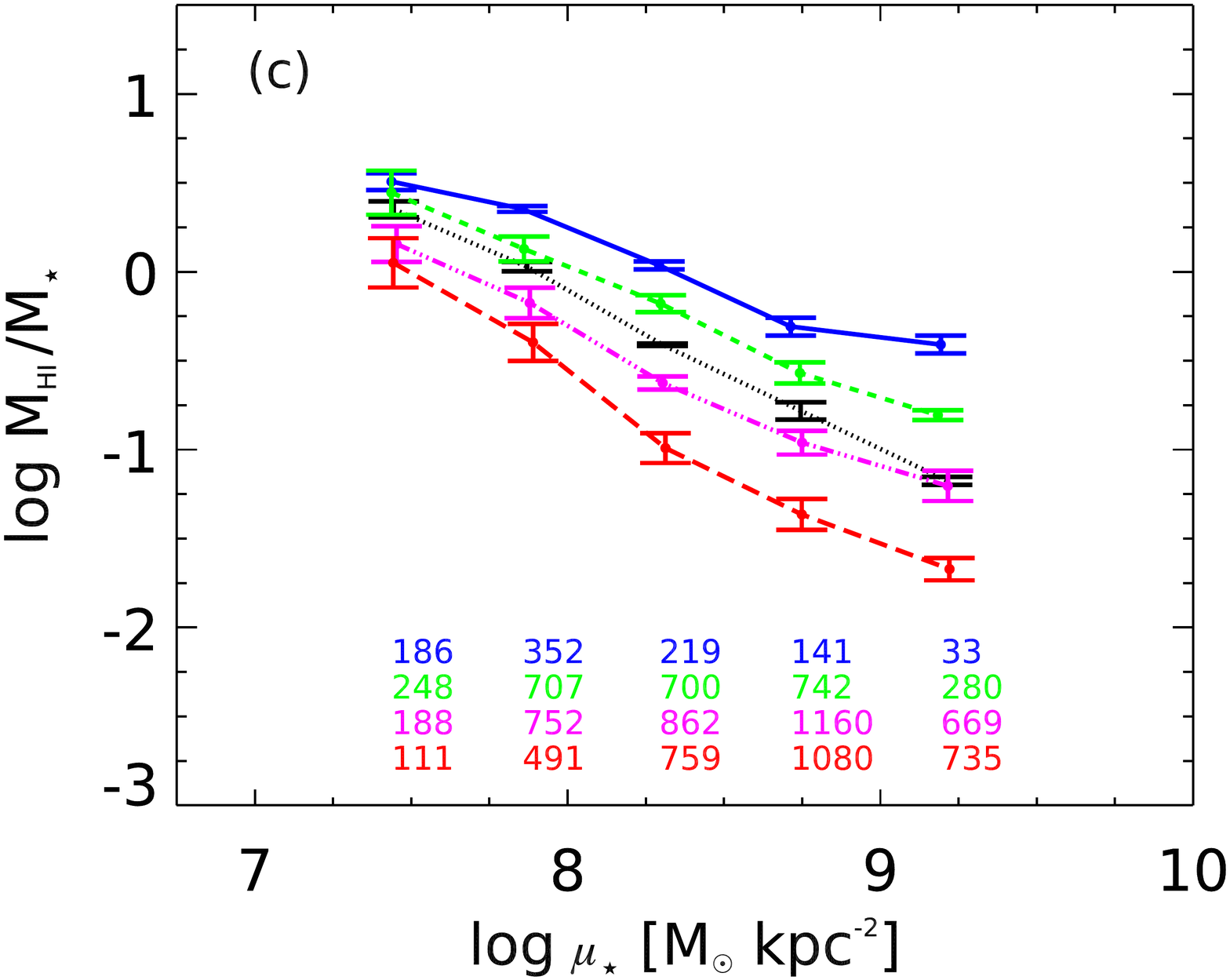}%
}
\end{center}
  \caption[LoF entry]{Average \HI gas fractions as a function of stellar mass (a), specific star formation rate (b) and stellar surface density (c) for \squiggle10,600 satellite galaxies. Black dotted lines show each scaling relation for the full sample of satellites. Solid blue, dashed green, dot-dashed magenta and long dashed red lines denote the relations when binned according to the halo mass of each satellite's host central. Halo mass limits are given in the legend and numbers along the bottom correspond to the sample statistics in each bin.}
\label{fig:main_relations}
\end{figure*}

In this section we present the main gas fraction scaling relations of \HInospace-to-stellar mass ratio versus stellar mass, sSFR and stellar surface density. This work disentangles, for the first time, the effects of mass, morphology and star formation on the gas content of satellite galaxies as a function of DM halo mass.

\begin{figure*}
\begin{multicols}{2}
    \includegraphics[width=\linewidth]{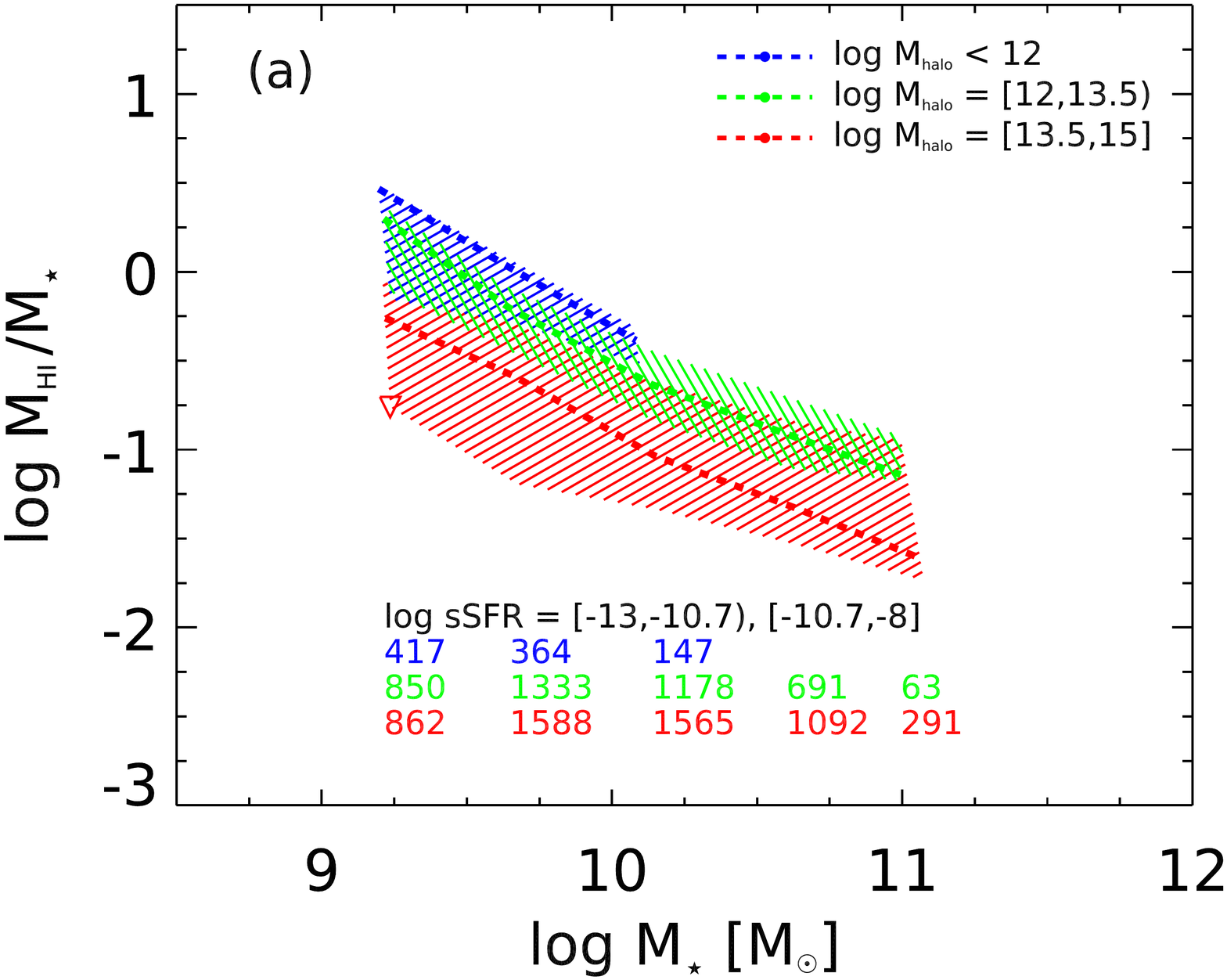}\par 
    \includegraphics[width=\linewidth]{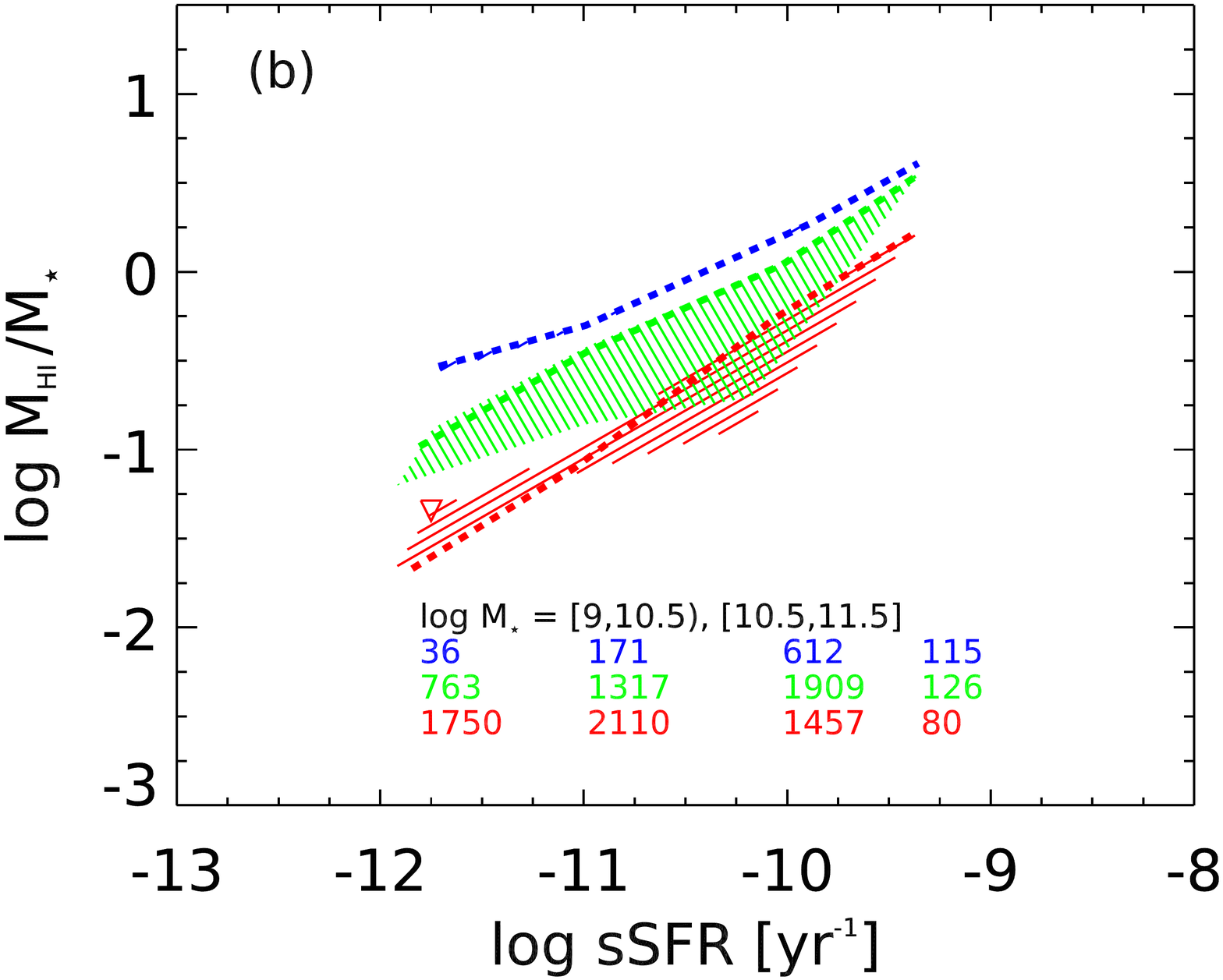}\par 
    \end{multicols}
  \caption[LoF entry]{Log \MHInospace/M$_{\star}$ versus log M$_{\star}$ (a) and log sSFR (b). The sample is binned according to the halo mass limits provided in the legend and plotted by the thick dashed lines. In addition, each halo mass cut in the left and right figures is also binned by sSFR and stellar mass respectively. On the left, the upper bound of each shaded region is given by the blue, star forming population (log sSFR/yr $\geq$ -10.7) while the red, quiescent galaxies (log sSFR/yr $<$ -10.7) provide the lower bounds. The right panel shows shaded regions for each halo mass bin bounded on the upper edge by galaxies in the low stellar mass bin (log \Mstarnospace/\Msol \space  $<$ 10.5) and lower edge by high mass galaxies (log \Mstarnospace/\Msol \space $\geq$ 10.5). Non-detections from stacking are included and are denoted by empty triangles. The numbers shown correspond to galaxies in each halo mass interval only, we do not show the number of galaxies in the stellar mass (b) and sSFR (a) envelopes. In order to ensure good statistics, we do not plot bins containing less than twenty satellites.}
\label{fig:mh_ms_shaded}
\end{figure*}

In Figure \ref{fig:main_relations}, we plot the stacked average \HI fraction as a function of stellar mass (a), sSFR (b) and stellar surface density (c; defined as \mustar = M$_{\star}$/2$\pi {\rm R}^{2}_{50,z}$ where R$_{50,z}$ is the Petrosian radius containing fifty per cent of the total z-band light in kpc) for satellite galaxies. The scaling relations for all satellites, not binned by environment, are denoted by the black dashed lines. Coloured lines show the stacked gas fraction relations in each of the halo mass bins given in the legend. At the bottom of each plot we provide the corresponding numbers of galaxies in each bin. Our chosen halo mass intervals divide the galaxies among the environments that are outlined in Table \ref{tab:env_class}. In the bin containing galaxy pairs, the question of whether such systems can be considered bound is a valid one. While interlopers and chance superpositions do occur, comparisons of the Y07 catalogue with detailed mock galaxy redshift surveys shows that the group finding algorithm performs remarkably well in this regime \citep[$>$95 per cent completeness; see][]{Yang2007}. Note that this work does not investigate the effect of environment in the small group regime, instead using it simply as the `zero-point' against which we compare the larger groups.

Figure \ref{fig:main_relations}a shows gas fraction versus stellar mass, separating the sample into the four environment bins. Note that there are no high stellar mass satellites in small groups, this is due to the abundance matching technique used to assign halo masses (see Section \ref{sec:FoF}). When the sample is split by halo mass we find that {\it at fixed stellar mass} there is a smooth and systematic reduction of satellite \HI content as halo mass increases. A satellite of log \Mstarnospace/\Msol \space = 10 which resides in a pair or small group is, on average, 0.2 - 0.5 dex more gas-rich than its stellar mass equivalent in a medium or large group, and has a gas fraction 0.8 dex larger than its cluster counterpart in a halo of log M$_{\rm h}$/\Msol \space $\geq$ 14. 

In Figure \ref{fig:main_relations}b sSFR is held constant and the sample is again separated by halo mass. In the two bins where log M$_{\rm h}$/\Msol \space $<$ 13, small to medium sized groups, the difference between average gas fraction across the range of sSFR is not significant (0.1 dex). {\it At fixed sSFR}, the galaxies in these haloes are statistically comparable in their average gas content. However, we do see large decreases (0.5 dex) in the average \HI fraction as a function of environment for galaxies with equivalent sSFR in haloes of log M$_{\rm h}$/\Msol \space $\geq$ 13. Not surprisingly, it is in the cluster regime where halo masses exceed log M$_{\rm h}$/\Msol \space $=$ 14 that we see the greatest impact (0.8 dex) on the \HI of satellites. 

We examine how \HI content varies as a function of halo mass at fixed stellar surface density in Figure \ref{fig:main_relations}c. Galaxies of a given surface density exhibit a large spread (0.9 dex) in gas fraction, with a smooth progression to lower \HI fraction across the range of environments. Note that the dispersion in gas content with halo mass at fixed surface density increases from small (0.5 dex) at disk-dominated, low densities to large (1.3 dex) at bulge-dominated, high densities. This is the result of the increasing contributions from the bulge in the measurement of R$_{50,z}$, meaning that the high surface density regime includes a fraction of galaxies that, while bulge dominated, still have a disk component.

Paper I established that, despite a residual dependence, stellar mass is not in fact an ideal tracer of neutral atomic hydrogen content and that sSFR is more closely related to the \HInospace-to-stellar mass ratio. Along with \HI, these two properties in particular correlate strongly with halo mass \citep[see][]{Wetzel2012,Wetzel2013}, thus when taking this analysis further one must check if decreases in \HI content as a function of halo mass are due to the sensitivity of gas to the external environment, or a consequence of stellar mass or sSFR properties. The large number of galaxies available allow us to disentangle these dual effects by controlling halo mass, sSFR and satellite mass simultaneously.

Figure \ref{fig:mh_ms_shaded} shows \HI fraction versus stellar mass (left panel) and sSFR (right panel) where the dashed coloured lines represent the average stacked gas fraction in each of the halo mass bins indicated. Note that, in order to increase statistics when selecting by many properties, it is necessary to reduce the number of halo mass bins from four to three. Again, we provide the number of galaxies in the corresponding bin along the bottom of each plot. 

In the left panel, we plot the gas fraction-stellar mass relation as function of halo mass, shading between bins of sSFR for each environment. The upper bounds of the shaded regions trace the average gas fraction as a function of stellar and halo mass for galaxies with log sSFR/yr $\geq -10.7$, whereas those systems with log sSFR/yr $< -10.7$ form the lower bound. This clearly shows that the residual dependence (0.5 dex on average) upon sSFR remains even when controlling for stellar and halo mass. Non-detections from stacking are plotted at their upper limit and marked with an open inverted triangle. By looking at the either the upper or lower bounds of each shaded region in Figure \ref{fig:mh_ms_shaded}a we are comparing satellites at fixed stellar mass and sSFR as a function of environment. For a given satellite mass in the blue cloud and red sequence we still see a decrease (on average 0.2 dex and 0.5 dex respectively) in gas fraction between each environment.

We apply a similar analysis to Figure \ref{fig:mh_ms_shaded}b. The dotted lines trace the gas fraction-sSFR relationship in each environment and we split each halo mass interval into low (log \Mstarnospace/\Msol \space $<$ 10.5) and high (log \Mstarnospace/\Msol \space $\geq$ 10.5) stellar mass satellites. The high and low mass relations form the outline of each shaded region on the top and bottom edges respectively. There is no blue shaded polygon because there are no galaxies in our sample that have a host halo mass of log M$_{\rm h}$/\Msol \space $<$ 12 and a stellar mass of log \Mstarnospace/\Msol \space $\geq 10.5$ (see Figure \ref{fig:main_relations}a). There is a scatter (0.4 dex on average) introduced to the gas fraction-sSFR relation in haloes above log M$_{\rm h}$/\Msol \space $=$ 12 by the residual effect of stellar mass on gas content. Despite the dependency on stellar mass, the second-order effect of environment on \HI shown in Figure \ref{fig:main_relations}b remains. Low and high mass satellites in more massive halos are gas poor (on average 0.3 dex and 0.5 dex respectively) at fixed sSFR compared to their counterparts in less massive halos.

\begin{figure}
      \centering
     \includegraphics[width=\linewidth]{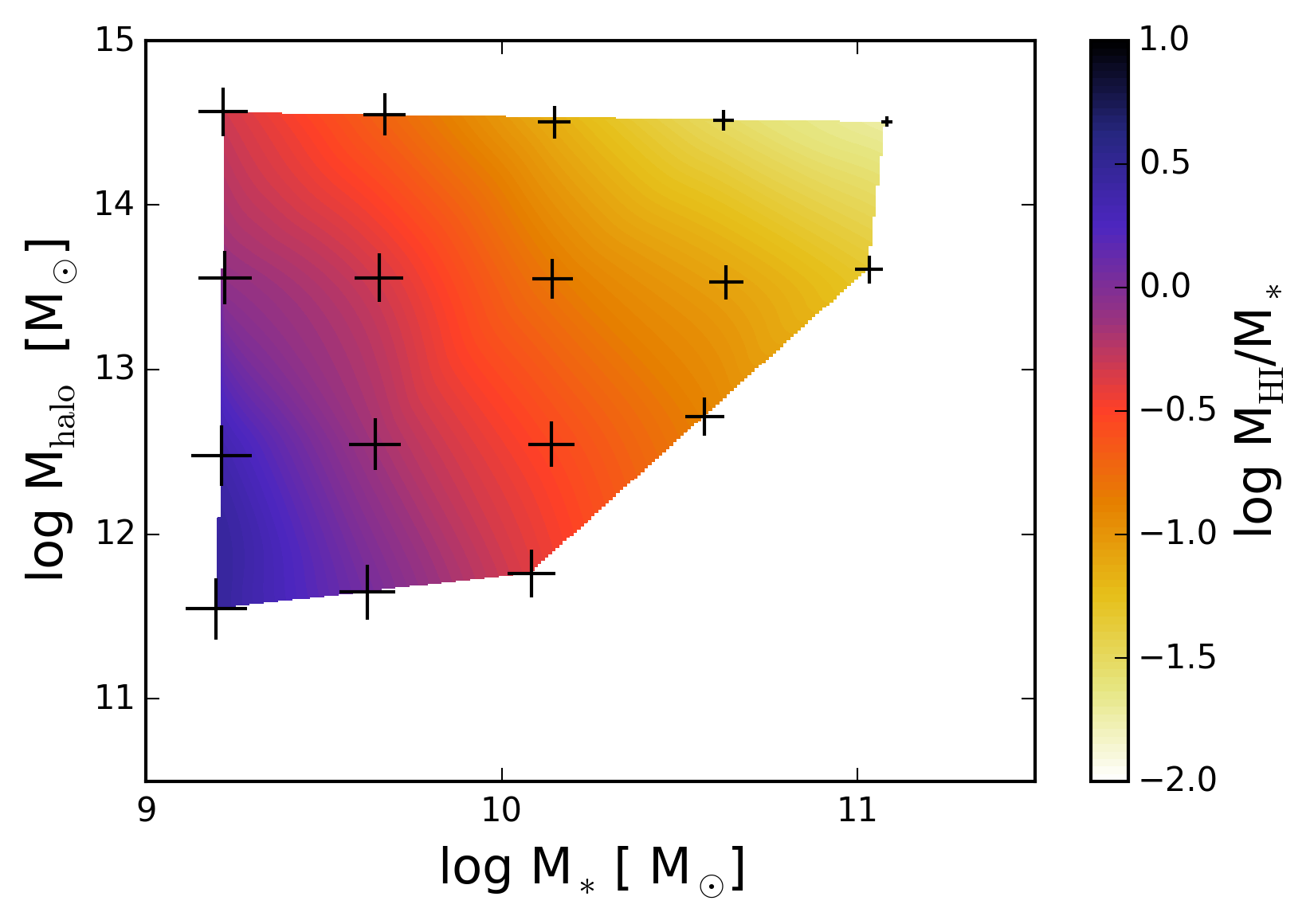}\par
  \caption[LoF entry]{Halo mass as a function of stellar mass for satellite galaxies, colour-coded by average gas fraction. There is a co-dependence of gas fraction on both environment and stellar mass. Black crosses are plotted at the mean halo and stellar mass values within each bin. The size of each marker is scaled to the gas fraction at that point.}
\label{fig:ms_mh_gf}
\end{figure}

We neatly show the simultaneous effects of stellar mass and environment on gas content in Figure \ref{fig:ms_mh_gf}, plotting the average gas fraction as a function of both host halo mass and satellite stellar mass. For illustration, we interpolate values between the bin centres marked by the black crosses. Contour colours reflect the average \HI fraction in that region of parameter space within our sample. The diagonal gradient of the colour change from gas-rich (purple) to -poor (yellow) shows the differential effects of stellar mass and halo mass upon the gas reservoirs of satellite galaxies.

Taken together, these results display the effect of halo mass upon the gas content of satellite galaxies as a function of stellar mass, sSFR and surface density. By holding constant mass and sSFR in Figure \ref{fig:mh_ms_shaded}, two galaxy properties shown to have a direct relationship with \HI content, we show that there is a strong secondary dependence of gas fraction upon environment across the group regime and into the cluster. In determining this, we break the degeneracy between internal processes that consume gas reservoirs, and external mechanisms which hinder the replenishment or encourage removal of gas from satellites. If satellite gas content is subject to physical mechanisms of an external origin (e.g. hydrodynamical pressure within the halo, gravitational interaction), the offset to lower gas fractions at fixed sSFR for halo masses above log M$_{\rm h}$/\Msol \space $=$ 13 suggests that \HI loss is occurring more quickly than the resulting quenching of star formation in these environments.

\section{The Influence of Local Density upon Gas Fraction}
\label{sec:LD_scalingRelations}

\begin{figure*}
\begin{multicols}{2}
     {\hspace{0cm}
          \textbf{7th Nearest Neighbour}\par}
	\includegraphics[width=\linewidth]{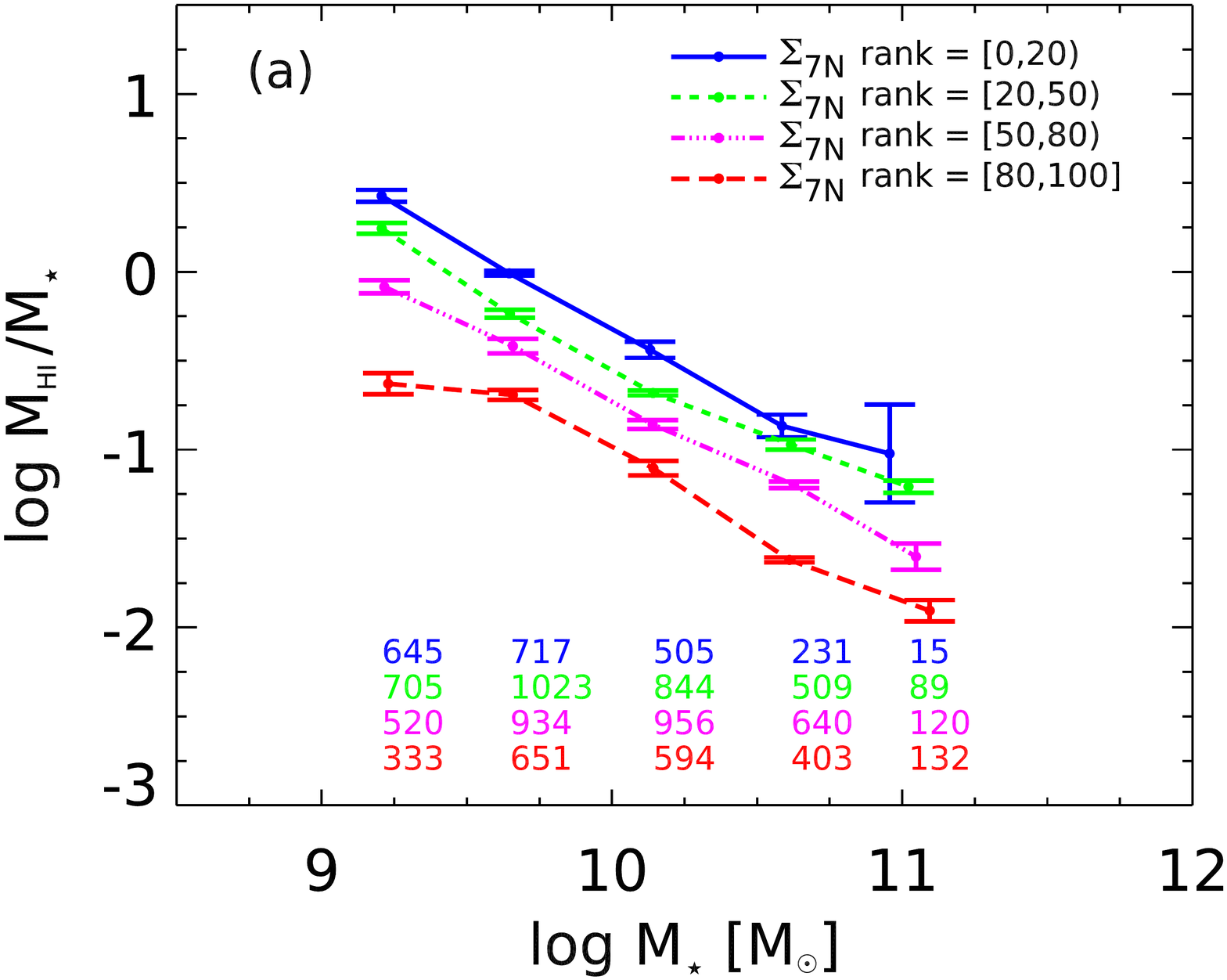}\par
     {\hspace{0cm}
          \textbf{Fixed Aperture}\par}
	\includegraphics[width=\linewidth]{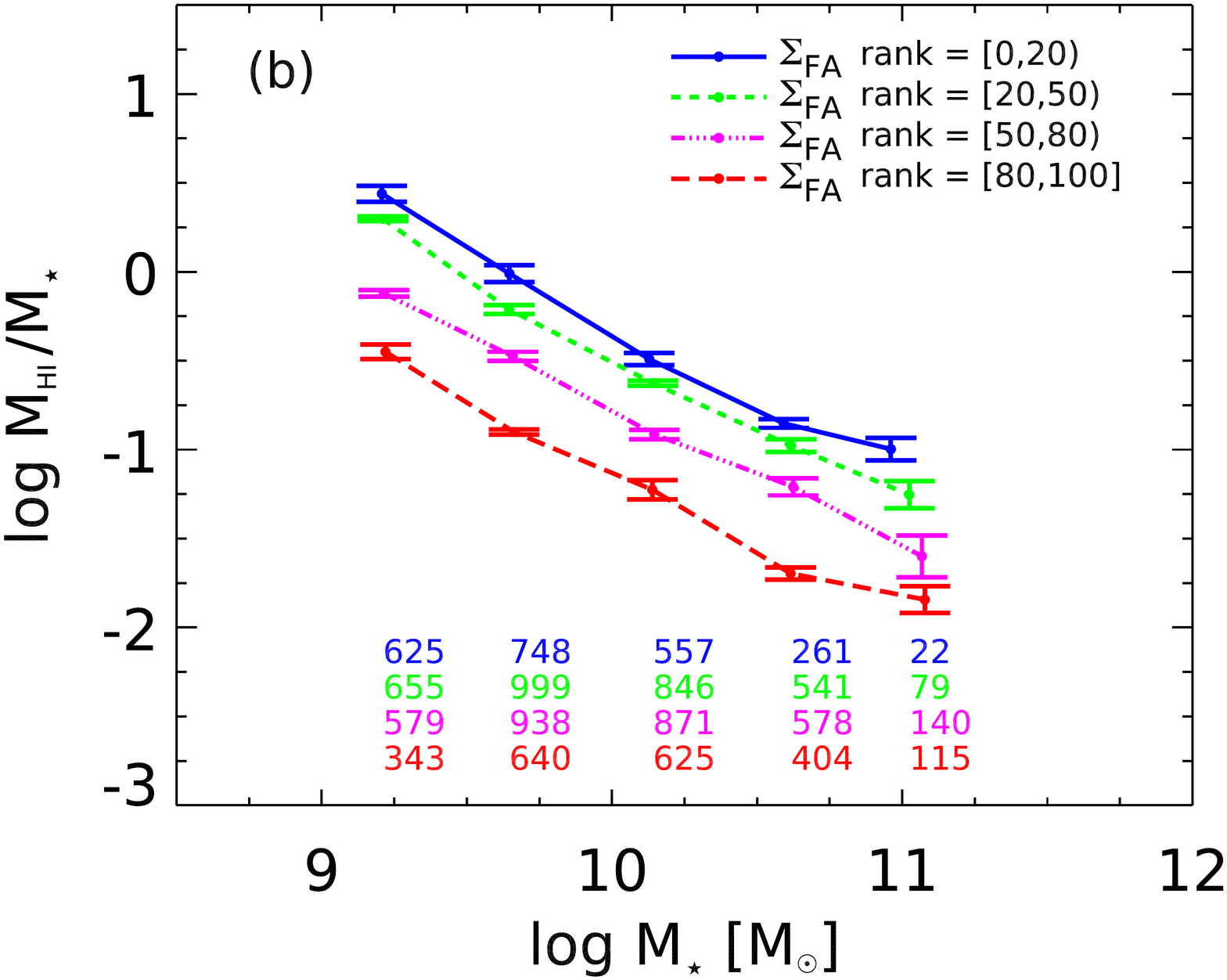}\par
    \end{multicols}
    \vspace{-1.6em}
\begin{multicols}{2}
	\includegraphics[width=\linewidth]{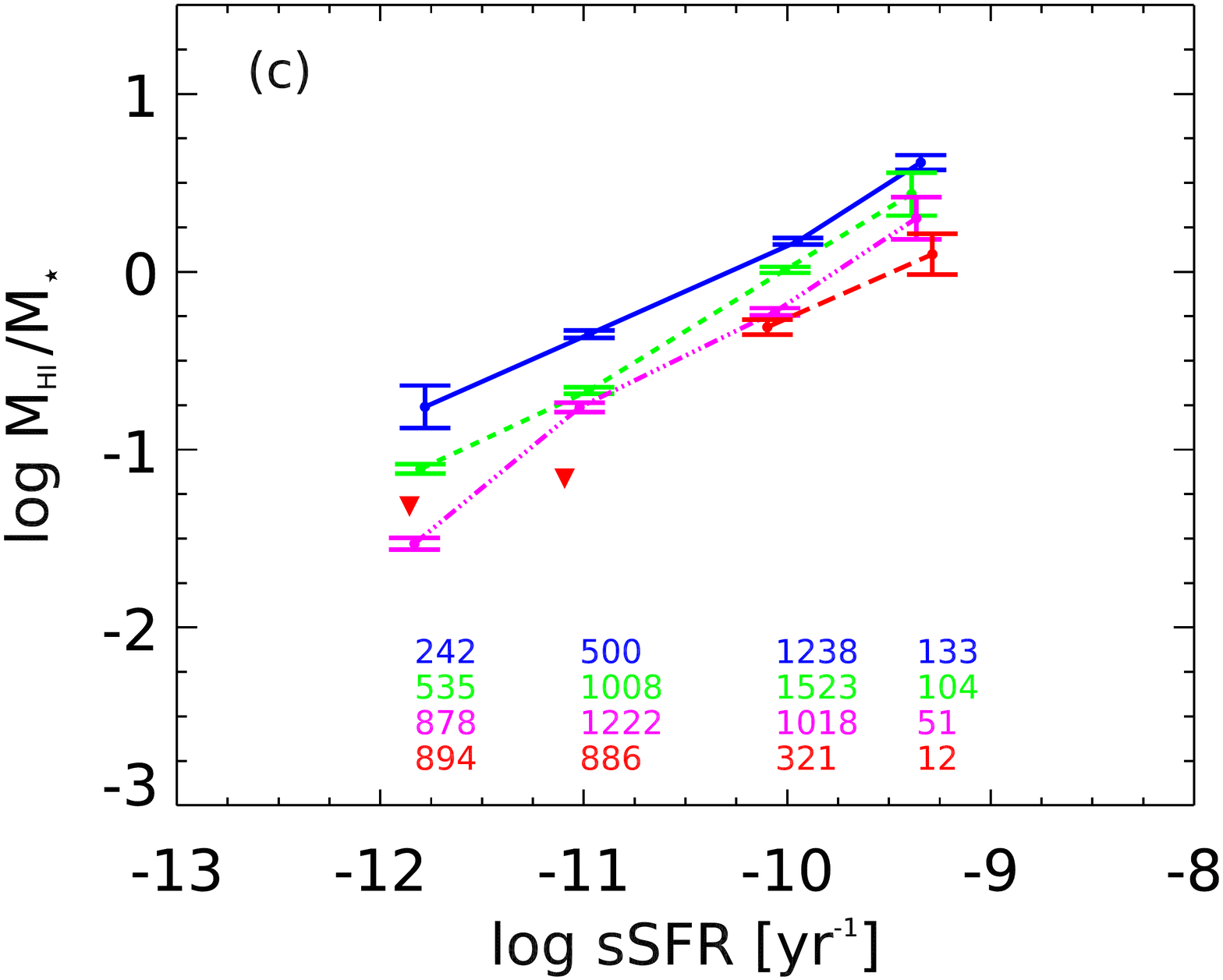}\par
	\includegraphics[width=\linewidth]{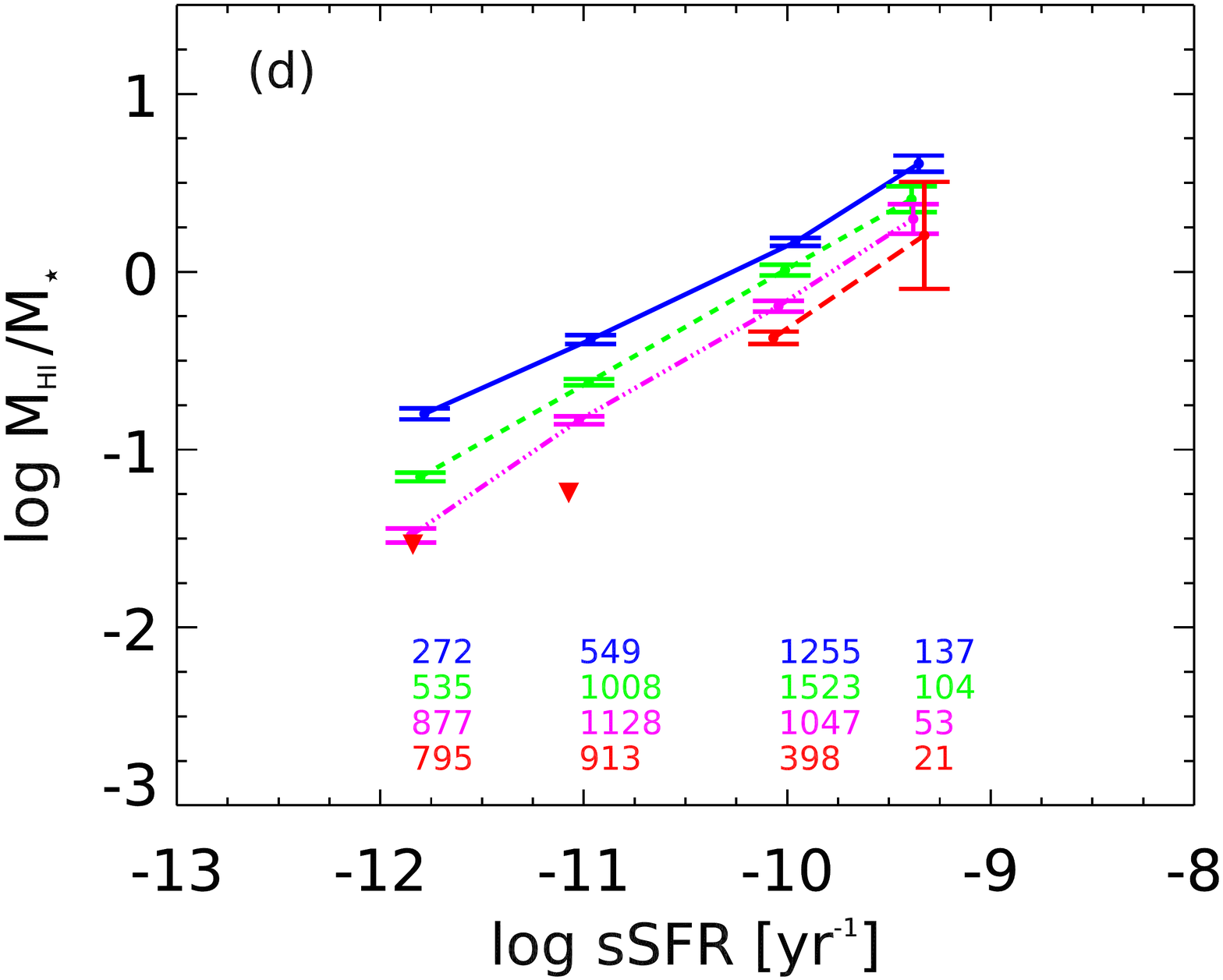}\par
\end{multicols}
  \caption[LoF entry]{On the top, (a) and (b) show log \MHInospace/\Mstar versus log \Mstar for satellite galaxies. On the bottom, (c) and (d) show log \MHInospace/M$_{\star}$ versus log sSFR. In the left panels (a, c), data are binned by the percentage rank of their 7th nearest neighbour density. The right panels (b, d) show satellites binned according according to their fixed aperture percentage rank. Bin limits are provided in the legends and numbers denote the statistics in each bin and non-detections from stacking are shown as upside-down triangles.}
  \label{fig:fa_sat}
\end{figure*}

Another way to characterise environments within a galaxy population is to use local density metrics. In this section we employ nearest neighbour and fixed aperture estimators in order to understand if our `definition' of environment is important when determining which environments and processes are the main culprits of the gas depletion seen in satellite galaxies.

Differences in the methods and distributions of halo mass, nearest neighbour and fixed aperture techniques can make direct comparison difficult. Therefore we convert the indicators for each galaxy to a {\it percentage rank}. To do so we rank the satellites in terms of each metric and assign percentages based upon the orders. For example, a galaxy with a nearest neighbour percentage rank of seventy-five will reside in an environment more dense than 75 per cent of other satellites and less dense than 25 per cent. This method enables us to compare the relative rather than absolute values of each environment metric.

\begin{figure*}
\begin{multicols}{2}
    \includegraphics[width=0.935\linewidth,valign=t]{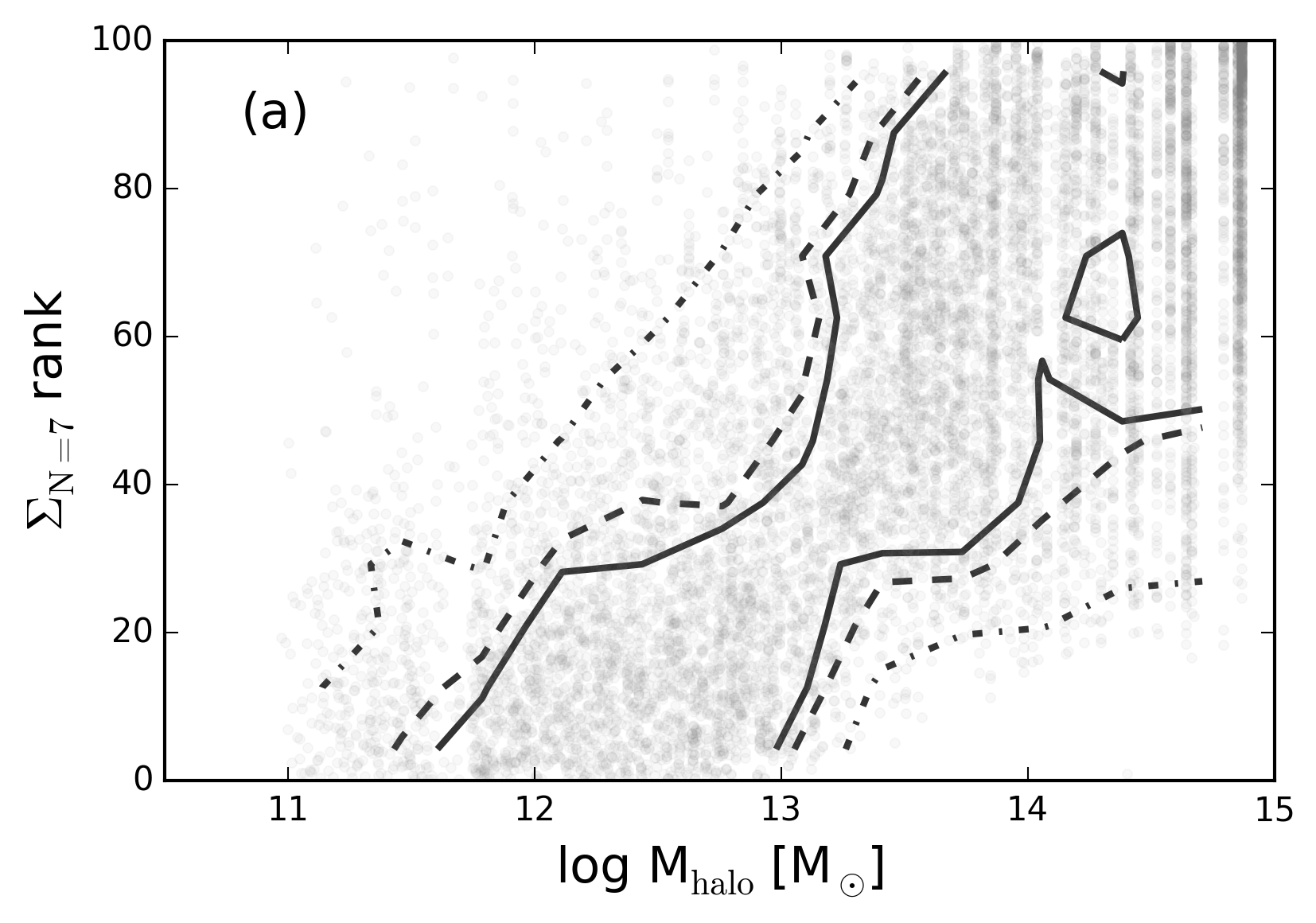}\par
    \includegraphics[width=0.935\linewidth,valign=t]{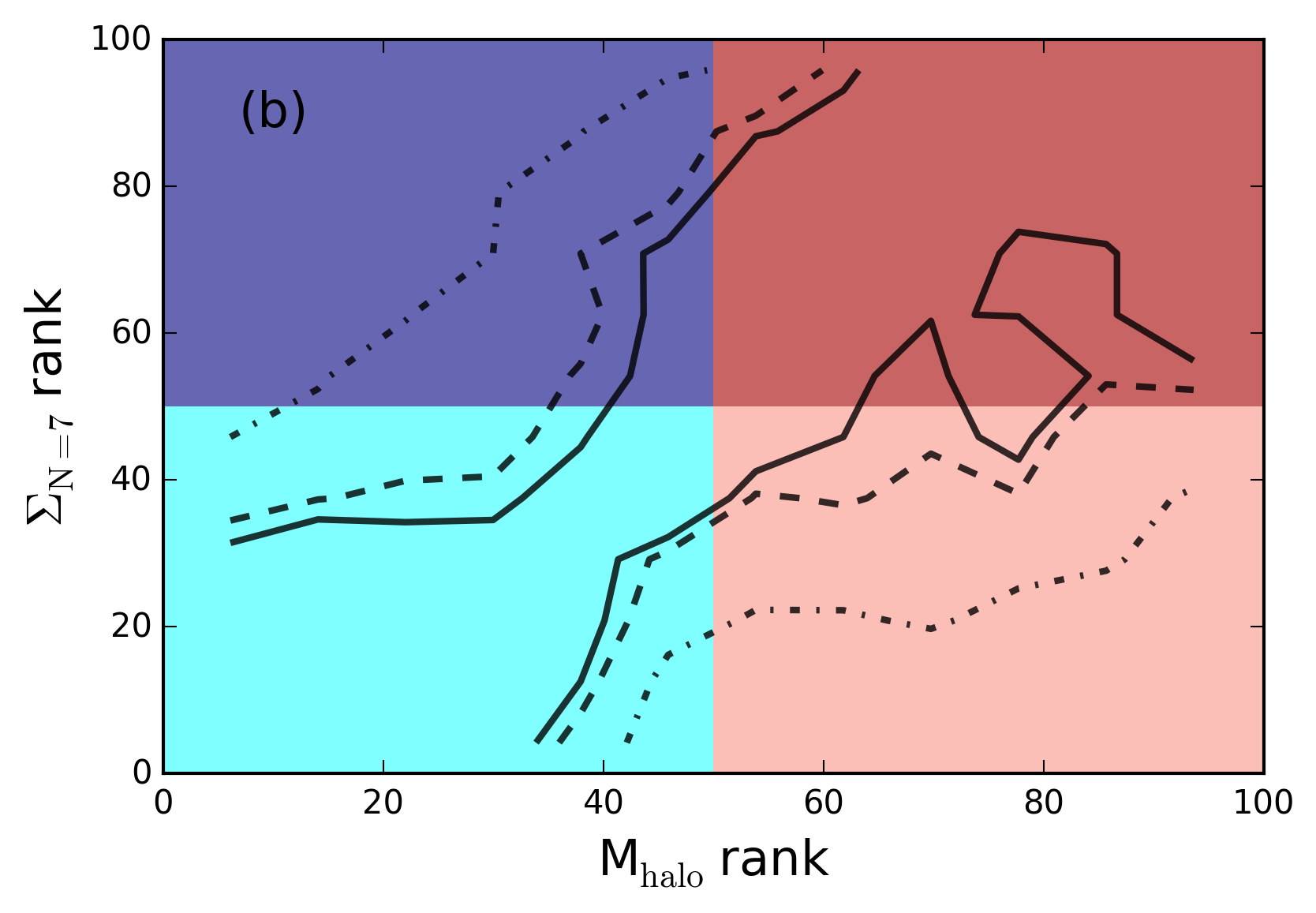}\par
    \end{multicols}
\begin{multicols}{2}
    \includegraphics[width=\linewidth,trim={0 0 0 1cm}]{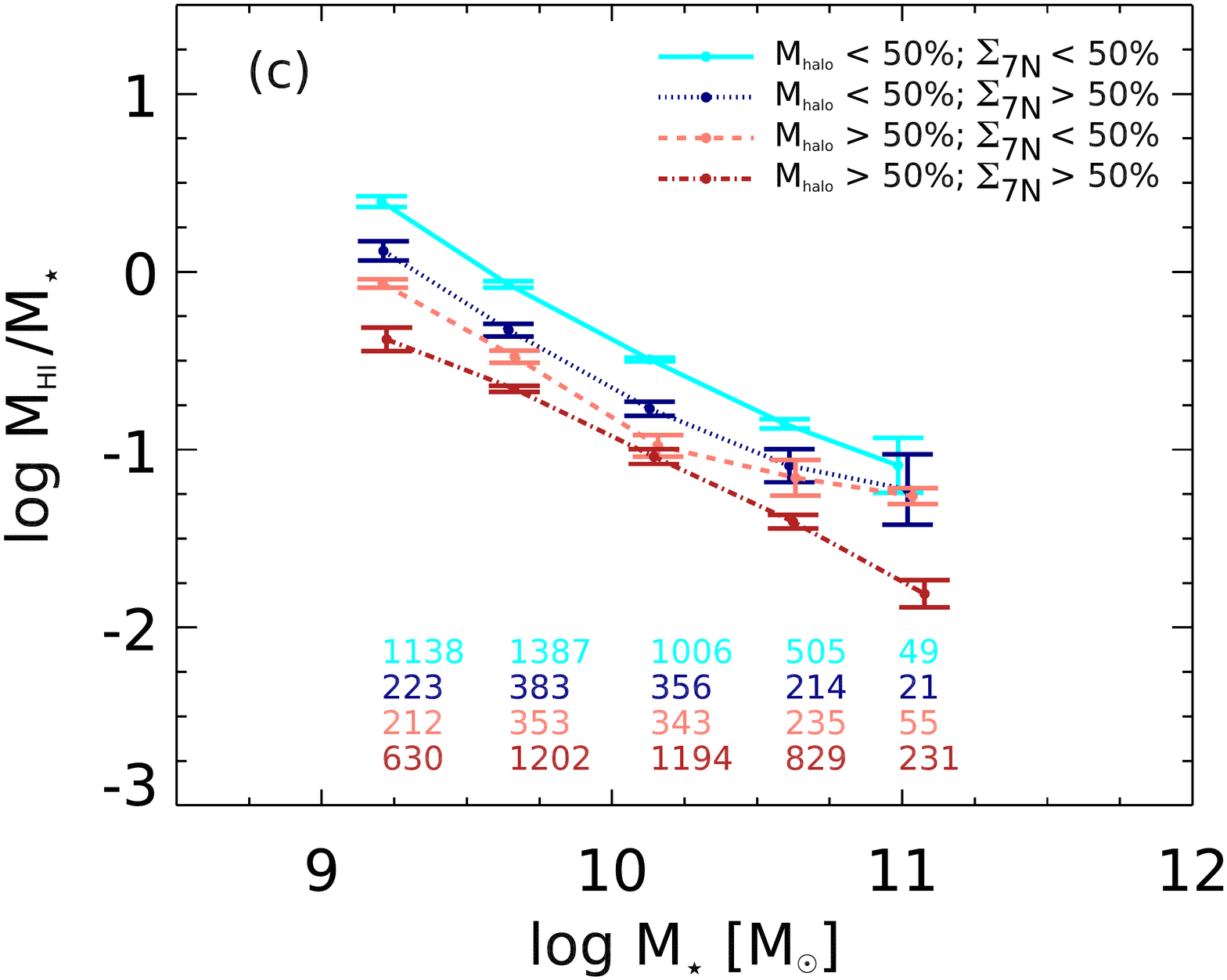}\par
    \includegraphics[width=\linewidth,trim={0 0 0 1cm}]{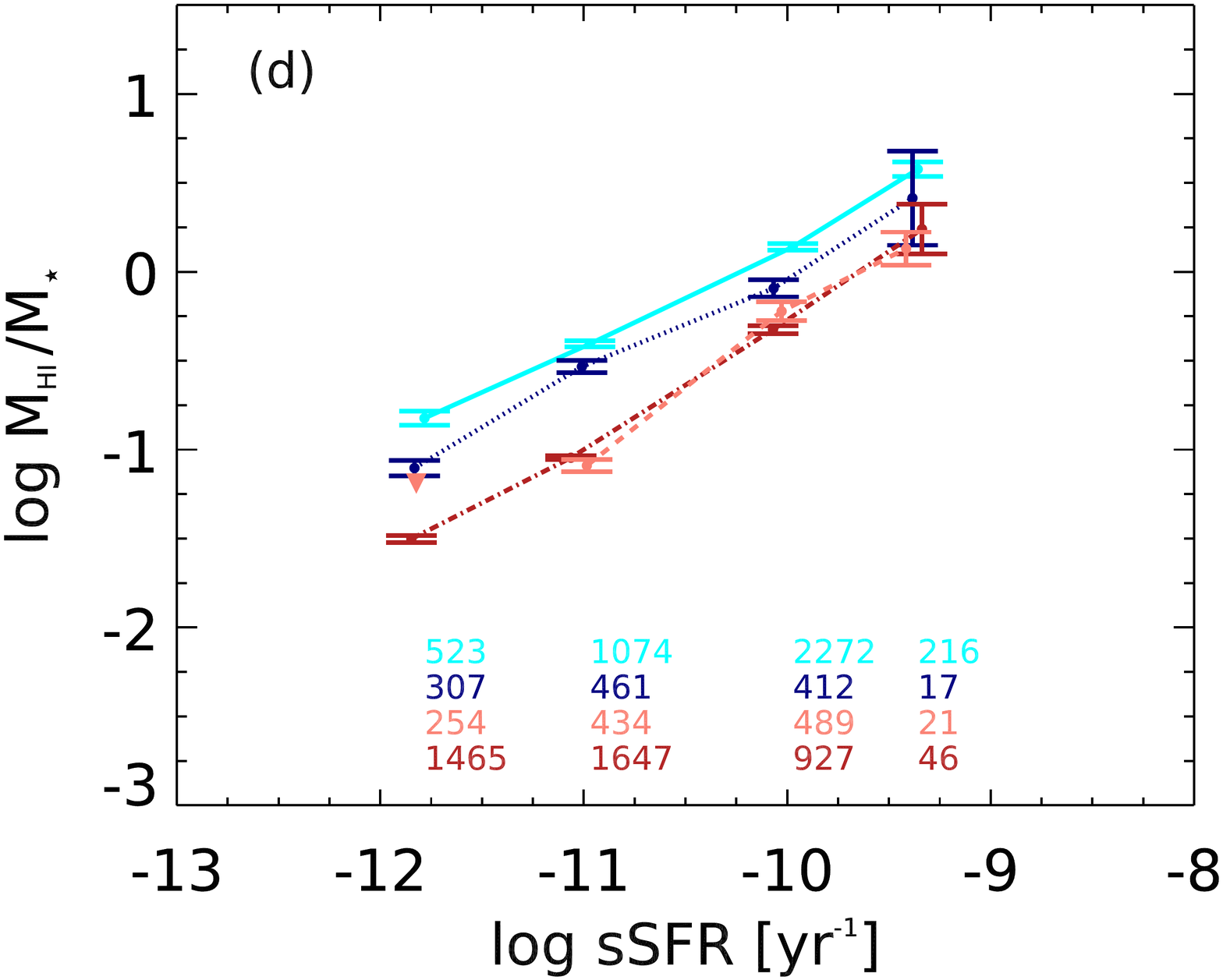}\par
\end{multicols}
  \caption{7th nearest neighbour density percentage rank of satellite galaxies as a function of halo mass (a) and 7th nearest neighbour density percentage rank versus log M$_{\rm h}$ percentage rank (b). Individual satellites in (a) are plotted in grey and contours are plotted at one (solid), two (dashed) and three (dotted) sigma levels in both panels. Log \MHInospace/\Mstar versus log \Mstarnospace, divided simultaneously into bins of halo mass percentage rank and 7th nearest neighbour percentage rank (c) and log \MHInospace/\Mstar versus log sSFR in the same bins (d). In these panels, the bounds of each bin are provided in the legend and the parameter space is illustrated by the corresponding colour panel in (b). Non-detections from stacked averages are denoted using upside-down triangles and the number of galaxies in each bin is shown along the bottom.}
  \label{fig:mh_n7_sat}
\end{figure*}

The two gas fraction scaling relations with stellar mass and sSFR, as function of local density percentage rank, are shown in Figure \ref{fig:fa_sat}. On the left, the coloured lines denote the sample galaxies separated according to their 7th nearest neighbour density percentage rank. On the right, the sample density is calculated using to a fixed aperture of radius = 1 Mpc ($\pm$ 1000 km s$^{-1}$). Again, numbers shown provide the count of galaxies in the relative bins and stacked averages resulting in non-detections are denoted by upside down triangles.

Figures \ref{fig:fa_sat}a and \ref{fig:fa_sat}b show that, in both cases, satellites that reside in denser environments are significantly more gas poor at a given stellar mass than galaxies that are found in less dense regions. Between the sparsest and densest regions there is a steady progression (0.6 and 0.7 dex for nearest neighbour and fixed aperture respectively) from high to low average gas fractions. In Figures \ref{fig:fa_sat}c and \ref{fig:fa_sat}d  we show that gas fraction also varies as a function of nearest neighbour and fixed aperture densities at fixed sSFR. For galaxies at a given sSFR, there is significant decrease (0.6 dex) in \HI content with increasing bins of both nearest neighbour and fixed aperture density.

We now look to determine whether the suppression of gas in the denser regions occurs because of the increase in galaxy number density and therefore chance of interaction, or is the consequence of the correlation between local density and halo mass. Figure \ref{fig:mh_n7_sat}a is a contour density plot showing percentage rank of the 7th nearest neighbour density versus halo mass for the sample of satellite galaxies. Grey points are individual galaxies and contours are set at the one, two and three sigma levels. There is a clear correlation between local density rank and halo mass with denser regions preferentially populating higher halo masses and visa versa. The interdependency is shown further once halo mass is ranked in the same manner by Figure \ref{fig:mh_n7_sat}b. For reference, the 50th percentile of the ranked halo masses corresponds to log M$_{\rm h}$/\Msol \space \space \squiggle 13.5 in absolute value, while 20 and 80 are log M$_{\rm h}$/\Msol \space \squiggle 12.5 and log M$_{\rm h}$/\Msol \space \squiggle 14.5 respectively.

We divide the parameter space of Figure \ref{fig:mh_n7_sat}b into colour shaded quadrants that correspond to the coloured relations of gas fraction as function of stellar mass and sSFR in Figures \ref{fig:mh_n7_sat}c and \ref{fig:mh_n7_sat}d respectively. For example, galaxies that are found in the dark red quadrant are included in the dark red, dot-dashed scaling relations, residing in the densest half of the sample and above the 50th percentile for halo mass. Numbers along the bottom in Figures \ref{fig:mh_n7_sat}c and \ref{fig:mh_n7_sat}d are the statistics in each bin.

Figure \ref{fig:mh_n7_sat}c shows that for a given stellar mass, gas reduction goes with both changes in local density and halo mass (0.6 dex). However, when one fixes density and alters the halo mass, comparing the cyan relation with the salmon and navy with the dark red, differences are larger (0.4 dex) than when density is changed at fixed halo mass (0.25 dex, cyan-navy/salmon-dark red). Interestingly, in Figure \ref{fig:mh_n7_sat}d, the effect is more noticeable. {\it At fixed sSFR, it is clear that gas fraction preferentially decreases with halo mass rather than density.} The differences in gas content between bins of varying halo mass (cyan-salmon/navy-dark red) are large (0.5 dex), while there is less of a difference ($<$ 0.2 dex) when only density changed. Figure \ref{fig:mh_n7_sat} clearly shows that the observed environmental suppression of gas in galaxies at fixed stellar mass and sSFR is dominated by the halo mass in which they reside and not the density of neighbours.

As previously mentioned, we check the validity of these results and the sensitivity to the aperture used by performing the same study out to the 3rd, 5th and 10th nearest neighbour, and with fixed apertures of 1 Mpc ($\pm$ 500 km s$^{-1}$), 2 Mpc ($\pm$ 1000 km s$^{-1}$) and 2 Mpc ($\pm$ 500 km s$^{-1}$). Changing the aperture size does not significantly affect our results, because of this we do not show the additional figures.

\section{Comparison with models}
\label{sec:models}

Section \ref{sec:ScalingRelations} has shown that, at fixed stellar mass, surface density and sSFR, satellite galaxies in more massive haloes have lower average \HI content than satellites in less massive haloes. Further to this, in Section \ref{sec:LD_scalingRelations} we determine that there is also a trend to lower gas fractions with increasing nearest neighbour and fixed aperture density, at fixed stellar mass and sSFR. We show the changes in the observed \HI content of satellites as function of environment at fixed sSFR to be primarily driven by the mass of the associated DM halo rather than their local density. 

The results extend findings from previous studies, showing that lower \HI fractions are found in the cluster environment \citep{Giovanelli1985,Cortese2011}, adding to the picture by demonstrating the systematic transition from gas-rich to -poor that occurs within groups, well before galaxies reach the cluster. Despite this, open questions remain surrounding the dominant processes that are responsible for the observed differences in gas fraction between low, intermediate and high mass haloes. We now look to identify the `fundamental' physics that is driving the environmental dependence seen in the scaling relations by connecting our empirical results to theoretical work.


\begin{figure*}
\raggedright
\hspace*{0.05575\textwidth}\begin{tabularx}{0.935\textwidth}{XXX}
  \adjustbox{center}{\textbf{GP14}} &
  \adjustbox{center}{\textbf{GP14+GRP}} &
  \adjustbox{center}{\textbf{``ezw'' Hydro}}\\
\end{tabularx}

\vspace{-1.6em}

\begin{center}
\resizebox{\textwidth}{!}{%
\includegraphics[trim={0 0 1.3cm 0},clip, valign=t]{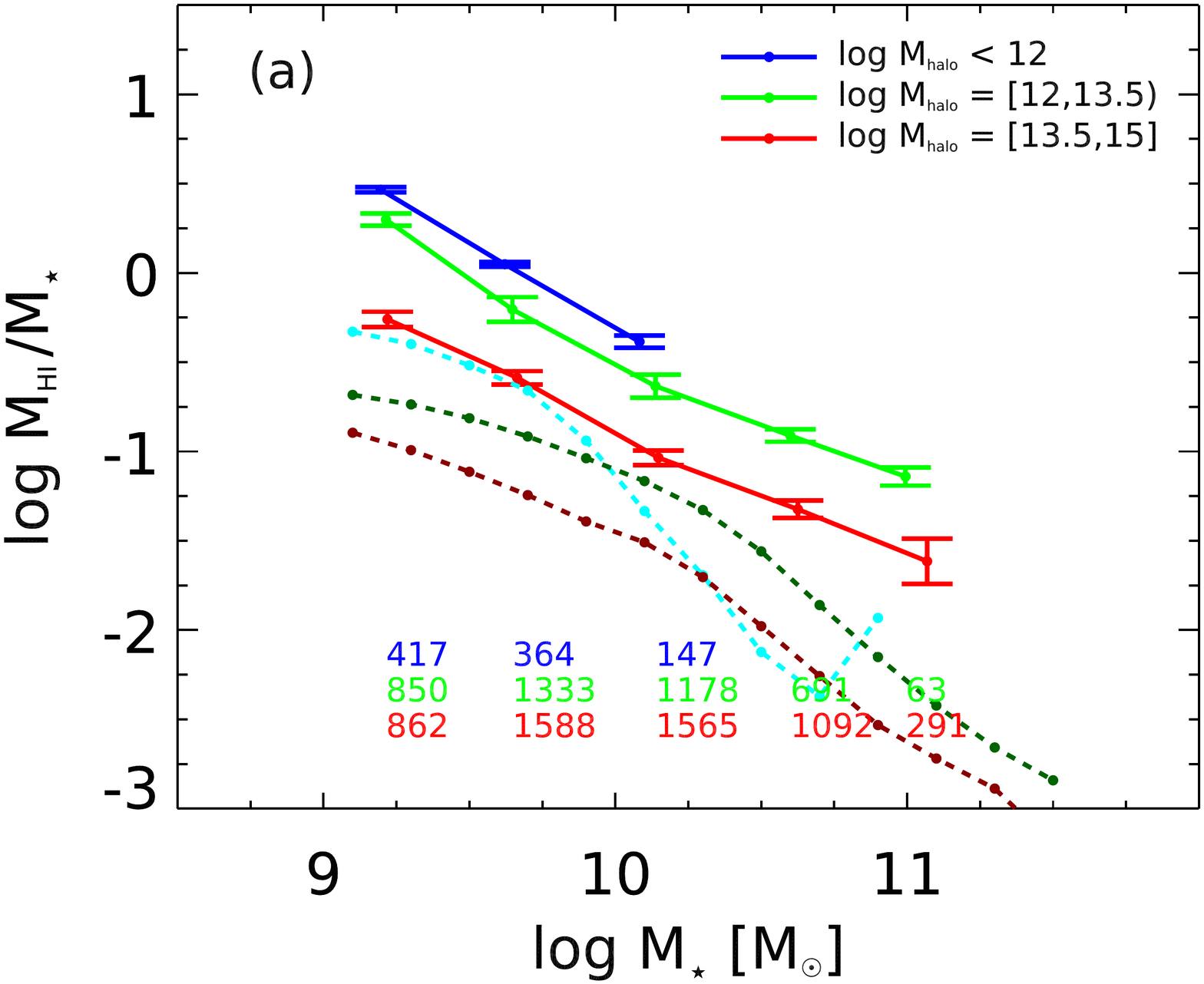}%
\hspace{-0.93cm}
\includegraphics[trim={3.8cm 0 1.3cm 0},clip, valign=t]{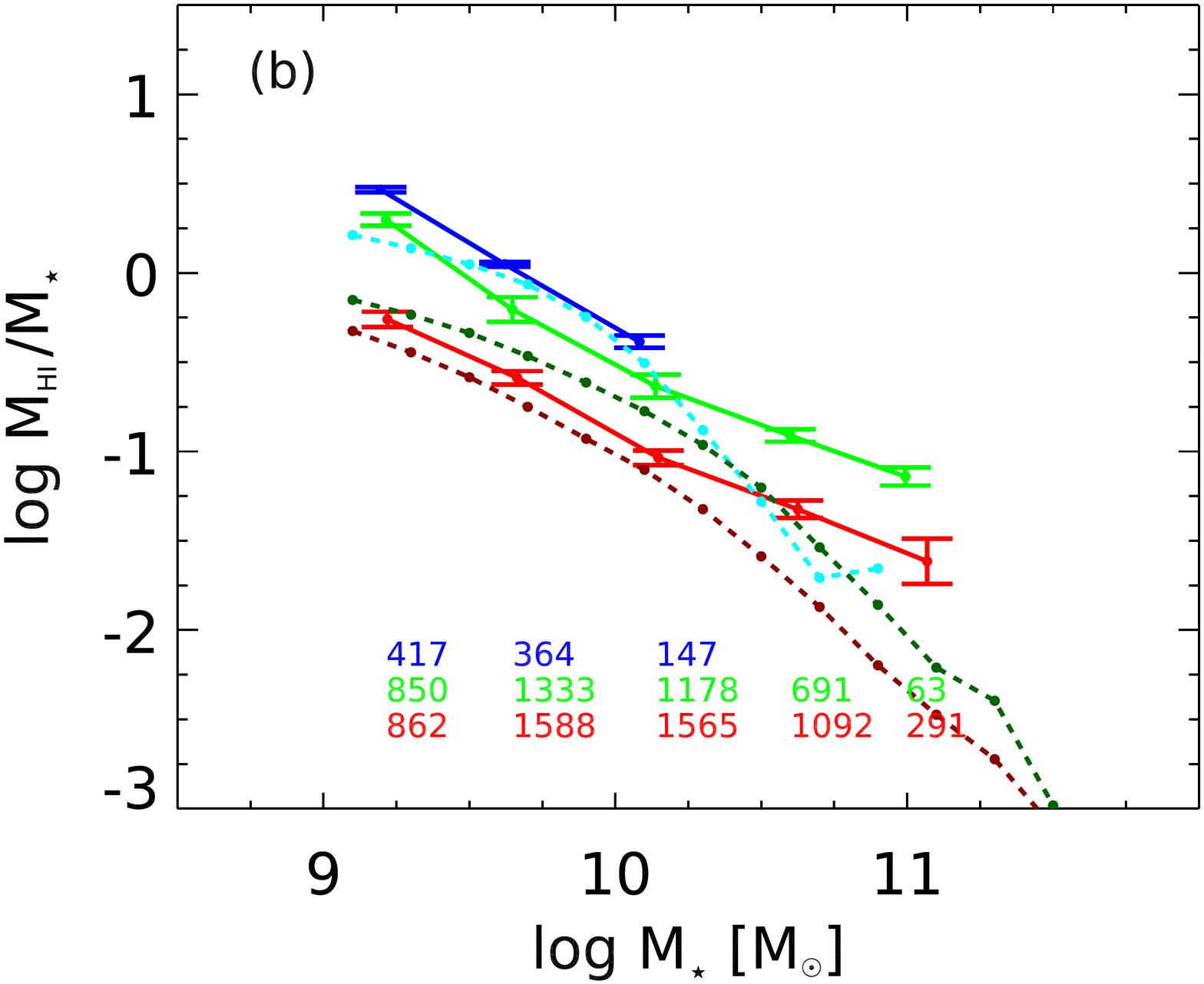}%
\hspace{-0.93cm}
\includegraphics[trim={3.8cm 0 1.3cm 0},clip, valign=t]{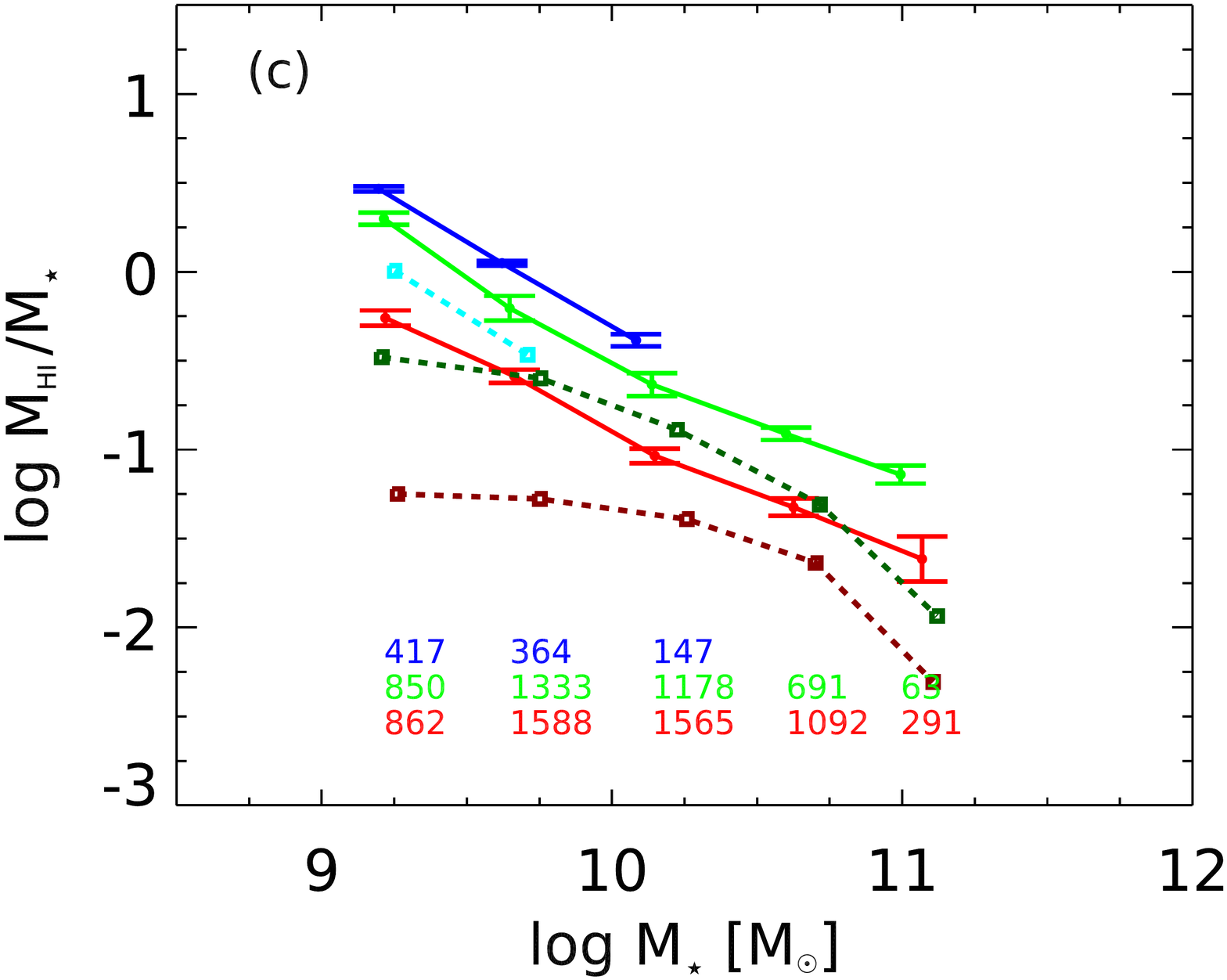}%
}
\end{center}
\vspace{-1.6em}
\begin{center}
\resizebox{\textwidth}{!}{%
\includegraphics[trim={0 0 1.3cm 0},clip, valign=t]{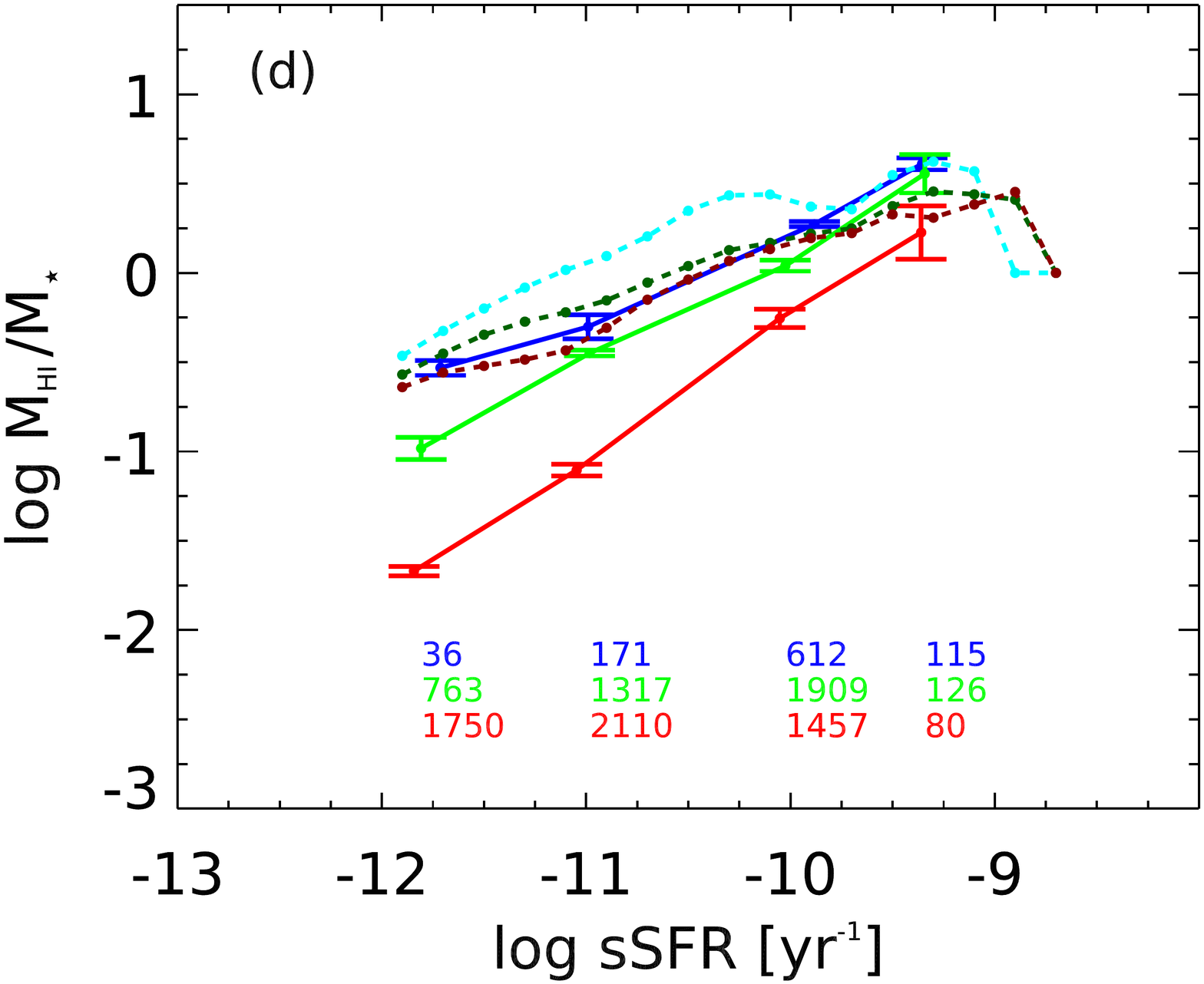}%
\hspace{-0.93cm}
\includegraphics[trim={3.8cm 0 1.3cm 0},clip, valign=t]{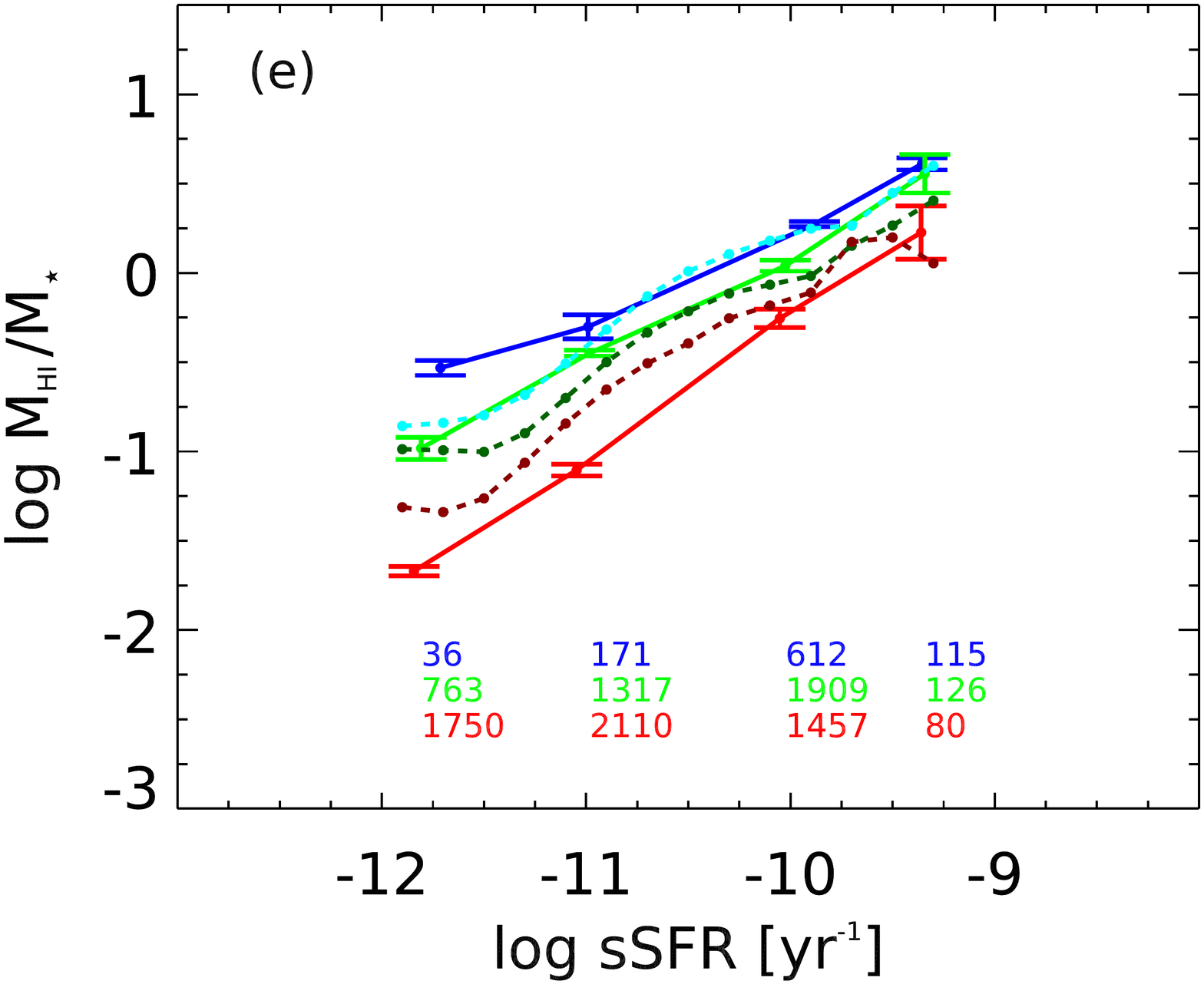}%
\hspace{-0.93cm}
\includegraphics[trim={3.8cm 0 1.3cm 0},clip, valign=t]{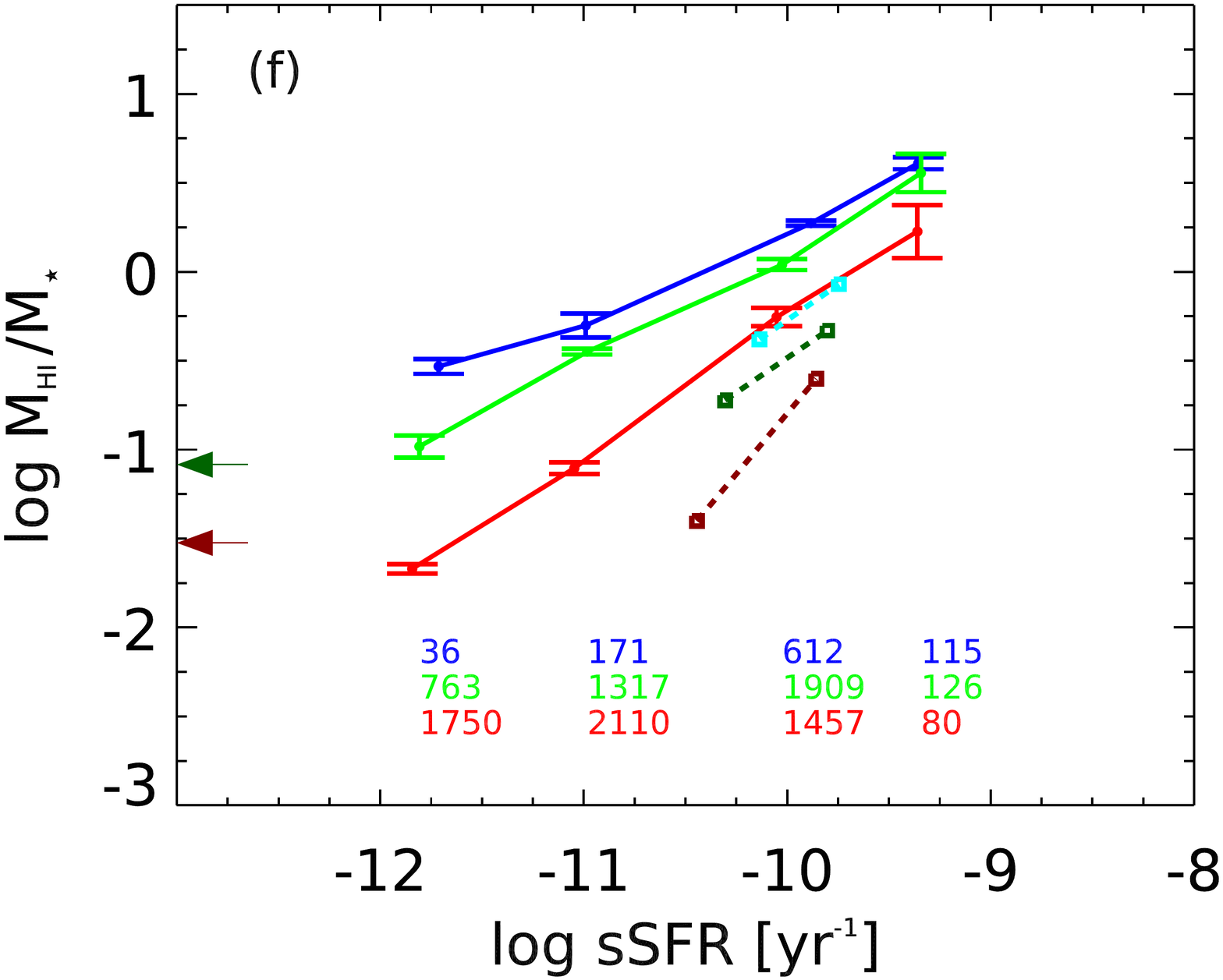}%
}
\end{center}
  \caption[LoF entry]{\HI fraction versus M$_{\star}$ (top) and sSFR (bottom) for satellite galaxies separated by halo mass. Solid lines are the same observations as the dashed lines shown in Figure \ref{fig:mh_ms_shaded}. Each column over plots a different galaxy formation model with cyan, dark green and dark red dashed lines corresponding to the same halo mass bins as the blue, green and red observational relations. Left: GP14 semi-analytic model, which assumes that once galaxies cross the virial radius of a larger halo they instantaneously lose their hot gas content and prevents further gas accretion. Middle: GP14+GRP model, which adopts gradual ram-pressure stripping of the hot gas and continued accretion of gas onto the galaxy. Right: Dashed lines show the ``ezw'' hydrodynamical simulation of \citet{Dave2013}, this models the stripping of the ISM from the disk. Arrows show upper limits on sSFRs set to the observational limit.}
  \label{fig:models}
\end{figure*}

The two most common approaches for conducting such a comparison are semi-analytic models and hydrodynamical simulations. In both cases, correctly modelling the influence of environment on the \HI content of galaxies is an extremely complex problem. Until relatively recently, much of the success in this area came from non-cosmological, high resolution simulations of well resolved galaxies \citep[e.g.][]{Marcolini2003,Mayer2006,Mayer2007,Bekki2009,Tonnesen2009,Tonnesen2010}. In recent years, the ability of cosmological models to reproduce observed trends in the global gas content of galaxies has improved significantly and successful comparisons have been made between theory and observations for general gas properties using both semi-analytic \citep[e.g.][]{Obreschkow2009,Power2010,Popping2014,Lagos2014} and hydrodynamical simulations \citep[e.g.][]{Dave2013,Rafieferantsoa2015,Crain2016,Marasco2016}. This success in replicating the global trends of gas content with key galaxy properties means it is important that we attempt understand how the models perform at reproducing the effect of environment upon the gas content of galaxies.

For our comparison, we choose one cosmological model from each camp; the semi-analytic model GALFORM \citep{Gonzalez2014} and the hydrodynamic simulations of \citet{Dave2013}, both have previously published successful comparisons with observational HI scaling relations \cite[see][]{Lagos2011b,Dave2013,Rafieferantsoa2015}. Below we describe the methodology of each and how they evolve the gas content in galaxies.

\subsection{Semi-analytic Models}

Semi-analytic models of galaxy formation typically treat each galaxy as a single object, using integrated properties and prescriptions to describe the baryonic physics governing their evolution. The primary advantage of this technique is its computational efficiency, allowing the production of statistical samples of galaxies that cover representative volumes and a large parameter space. One caveat is that bulge and disk properties (gas, stars, SFR etc.) are described by a single number for each component, meaning the internal dynamics and physics are not resolved.

In this subsection we compare our results with the semi-analytic simulation GALFORM \citep{Gonzalez2014}. The motivation for choosing this model is its environmental treatment of the hot gas and tracking of the cold gas as galaxies evolve.

\subsubsection{The GALFORM Model}

The GALFORM semi-analytic model includes the main physical processes that are considered to shape the formation and evolution of galaxies \citep{Cole2000}. These are: (i) the collapse and merging of DM haloes, (ii) the shock-heating and radiative cooling of gas inside DM halos (which lead to the formation of galactic disks), (iii) star formation in disks, (iv) feedback from supernovae, from active galactic nuclei and from photo-ionization of the inter-galactic medium, (v) chemical enrichment of stars and gas due to stellar evolution, (vi) galaxy mergers driven by dynamical friction (which trigger starbursts and lead to the formation of bulges), (vii) global disk instabilities (which also lead to the formation of bulges), and (viii) ram-pressure stripping of the hot gas. For this paper we focus on the published version of GALFORM of \citet{Gonzalez2014}, hereafter GP14.

In GP14, the halo merger trees are extracted from the updated version of the Millennium $N$-body simulation \citep{Springel2005b} using WMAP7 \citep{Komatsu2011}. \citet{Gonzalez2014} also includes the explicit tracking of the atomic and molecular cold gas component in galaxies \citep[by using the hydrostatic midplane pressure of disks and bulges as a proxy of the atomic-to-molecular gas surface density ratio;][]{Blitz2006,Leroy2008} as well as the hot gas phase. Star formation is characterised following \citet{Blitz2006}, who use an empirical  formulation of the Kennicutt-Schmidt law
\begin{flalign}
&\Sigma_{\rm SFR} = \tau_{\rm SF} \Sigma_{\rm H_{2}} &
\label{eq:SF_law}
\end{flalign}
where $\Sigma_{\rm SFR}$ is the star formation rate surface density, $\tau_{\rm SF}$ is the inverse star formation timescale and $\Sigma_{\rm H_{2}}$ is the surface density of molecular gas. The current implementation was developed by \citet{Lagos2011b}, further details can be found in that paper.

Here, we present results from two variants of the GP14 model which differ in their treatment of the hot gas of satellite galaxies once they cross the virial radius of the larger halo. The first variant (which we simply refer to as GP14 as it is the default implementation) assumes {\it instantaneous} stripping of the hot gas; once a galaxy becomes a satellite, its hot gas is removed and transferred to the gas reservoir of the main halo. By removing satellite hot gas, GP14 halts the replenishment of cold gas reserves through accretion, forcing each galaxy into a state of `strangulation'. The second variant, which we refer to as GP14+GRP, assumes instead {\it gradual} ram-pressure stripping of the hot gas, which in practice leads to satellite galaxies having accretion rates that continuously decay in time once they become satellite (as opposed to a sharp shut off of the accretion rate in the first variant).  GP14+GRP defines the `stripping radius' as the point at which ram-pressure and gravitational pressure are balanced, gas that resides beyond this radius is removed from the satellite. The ram and gravitational pressure prescriptions are detailed in \citet{Lagos2014} and follow the respective functional forms
\begin{flalign}
&{\rm P_{\rm ram}} = \rho_{\rm gas,p} v_{\rm sat}^2 &
\label{eq:rp_law}
\end{flalign}
and 
\begin{flalign}
&{\rm P_{\rm grav}} = \alpha_{\rm rp} \dfrac{G {\rm M}_{\rm tot,sat}(r_{\rm str}) \rho_{\rm gas,sat}(r_{\rm str})}{r_{\rm str}} &
\label{eq:gravp_law}
\end{flalign}
where ${\rm P_{\rm ram}}$ is the ram-pressure, $\rho_{\rm gas,p}$ is the gas density of the parent halo and $v_{\rm sat}$ is the satellite's velocity with respect to its parent halo. ${\rm P_{\rm grav}}$ is the gravitational pressure, $\alpha_{\rm rp}$ is a geometric constant of order unity and set to 2, $G$ is the gravitational constant, ${\rm M}_{\rm tot,sat}(r_{\rm str}) $ is satellite's total mass (including stars, gas and DM) within the stripping radius $r_{\rm str}$ and $\rho_{\rm gas,sat}(r_{\rm str})$ is the satellite's hot gas density at  $r_{\rm str}$.

It is important to note that neither implementation of GP14 used in this work accounts for ram-pressure stripping of the cold gas, a process known to dominate in larger halos \citep[i.e. galaxy clusters;][]{Boselli2006,Tonnesen2007,McCarthy2008,Chung2009,Cortese2010,Cortese2011}. Other processes expected to drive mass loss in satellite galaxies (e.g. tidal stripping of stars, heating due to tidal shocks, harassment) are not included in either of the variants. In cases where the dynamical mass of the satellite is significantly below that of the group, such mechanisms have been shown to remove cold gas in significant amounts over many Gyrs, however, such effects are expected to be dominant in dwarf galaxies, i.e. those with circular velocities $\lesssim 30$ km s$^{-1}$, which are smaller than the galaxies we are studying here \citep[see][]{Mastropietro2005,Mayer2006,Tomozeiu2016}.

The GP14+RP variant has been shown to produce both atomic and molecular gas fractions of early-type galaxies, and fractions of passive galaxies as a function of stellar mass that are in better agreement with the observations \citep{Lagos2014,Guo2016}.

\subsection{Hydrodynamical Simulations}

Compared to semi-analytics, hydrodynamical simulations reproduce the processes that govern galaxy evolution at much higher resolution, solving the equations of gravity, dynamics and radiative transfer for up to 10$^6$ particles in each galaxy. With the caveat that mechanisms such as star formation, stellar feedback, black hole accretion still occur below the resolution limit, this approach has the advantage of galaxies being resolved into several elements and no imposed assumptions of how gas accretion takes place or the influence of DM on galaxy properties (e.g. sizes and accretion rates). These models have been successful in their reproduction of realistic objects, however, modelling such detailed physics on small scales is computationally very expensive, therefore sample size and parameter space explored is smaller than that of semi-analytics.

\subsubsection{\citet{Dave2013} ``ezw'' Model}
\label{sec:ezw_desc}
We compare to the cosmological hydrodynamic simulations of \citet{Dave2013}, who used a modified version of Gadget-2 \citep{Springel2005b} to study the \HI content of galaxies. The simulation uses a $\Lambda$CDM cosmology with $\Omega_{\rm m}=$ 0.28, $\Omega_{\Lambda}=$ 0.072, $\Omega_{\rm baryons}=$ 0.046, $\sigma_{8}=$ 0.81, $n_{s}=$ 0.96 and $h=$ 0.70. There are 512$^3$ DM particles and 512$^3$ gas particles in a cubical, comoving volume of 32 $h^{-1}$ Mpc on each side. The DM and gas particle resolution is 2.3 $\times$ 10$^7$ \Msol \space and 4.5 $\times$ 10$^6$ \Msol \space respectively. In the model halos and galaxies grow self-consistently from DM and gas particles. Here we briefly summarise the processes of galaxy formation and evolution modelled in the simulation: (i) primordial and metal line cooling based upon the photo-ionization equilibrium of \citet{Wiersma2009}, (ii) star formation following a Schmidt law \citep{Schmidt1959} where SFR is proportional to gas density, applied using the sub-grid recipe of \citet{Springel2003}, (iii) stellar and supernovae feedback provides metal enrichment following the prescriptions of \cite{Oppenheimer2008}, (iv) quenching energy comparable to AGN feedback is imparted on massive galaxies by heating infalling gas (to fifty times the virial temperature) once the halo mass, estimated from the individual galaxy mass, exceeds around log M$_{\rm h}$/\Msol \space = 12, (v) finally, strong galactic outflows (their ``ezw" model), driven through a hybrid of momentum flux from young stars and energy from supernovae, are assumed to kinetically eject gas from the ISM.

Each halo is identified using a spherical over-density based approach, while galaxies are identified using Spline Kernel Interpolative Denman \citep[SKID; see ][]{Dave2013}. For halos with multiple resolved galaxies (which is the majority for galaxies with log \Mstarnospace/\Msol \space $\geq$ 9), the largest stellar mass galaxy is identified as the central, and the others are satellites. \HI is computed within the model by determining the optically thin, neutral fraction of each gas particle. \citet{Dave2013} then separate the neutral gas into its atomic and molecular phases based upon the ISM pressure prescriptions of The \HI Nearby Galaxy Survey \citep[THINGS;][]{Leroy2008}.

Once a satellite enters the halo of another galaxy, the \HI may be influenced by the following environmental processes, each of which are modelled self-consistently within the simulation: ram-pressure and viscous stripping \citep{Marcolini2003}, tidal interaction and harassment, and strangulation of inflowing gas. For further details on how these mechanisms are implemented and their dependence on halo properties see \citet{Rafieferantsoa2015}. An important caveat to note is that \citet{Dave2013} employ entropy-conserving smoothed particle hydrodynamics (SPH) \citep{Springel2005b} that has been shown to handle surface instabilities such as those that occur during gas stripping poorly compared to more recent hydrodynamics methods \citep[see][]{Agertz2007,Hopkins2015,Schaller2015}. This issue and its possible implications are discussed further in Section \ref{sec:comp-obs}.

\subsection{Comparison with Observations}
\label{sec:comp-obs}
In Figure \ref{fig:models} we compare our observations of gas content at fixed stellar mass (top panels) and sSFR (bottom panels) with the models. The \HI gas content of satellite galaxies in the simulations is  calculated in a way that is identical to our stacking procedure (i.e. $\text{log} \: \langle \text{M}_{\text{HI}}/\text{M}_{\star} \rangle$). We split satellites according to identical bins of halo mass for both observations  (blue, green, red) and theory (cyan, dark green, dark red). The number of galaxies in each observational bin is plotted along the bottom.

We see that the GP14 model predicts satellites that are too gas poor at fixed stellar mass (\ref{fig:models}a) and too gas rich at fixed sSFR (\ref{fig:models}d). While this seems in contradiction, the process of instantaneously removing a galaxy's hot gas from its subhalo once it becomes a satellite leads to an overly quenched satellite population. Naturally, these systems have low or negligible amounts of cold gas which means that, where they are included in the scaling relations (i.e. \ref{fig:models}a), average gas fractions are artificially low. On the other hand, extremely quenched systems are removed from the gas fraction-sSFR plane by construction. Therefore they do not contribute to the average \HI content and one is not comparing the same GP14 satellite population between Figures \ref{fig:models}a and \ref{fig:models}d. Once this is added to the picture, the two plots are in agreement and it becomes clear that the stripping implemented by GP14 is too strong and too rapid to match observations.

In Figure \ref{fig:models}b we find that the GP14+GRP model, which assumes the hot gas gets stripped gradually as satellite galaxies travel through their host haloes, is in better agreement with the observations for stellar masses of log \Mstarnospace/\Msol \space $\lesssim$10.3 in the highest and lowest halo mass bins studied. This is because in the GP14+GRP model, satellite galaxies are able to retain their hot gas for a longer timescales, which in turn means that they are able to replenish their ISM for longer. Nevertheless, at these stellar masses, the GP14+GRP model is still on average \squiggle 0.2 dex too low compared to observations at fixed stellar mass. This is because the models predict sSFRs that are usually lower than the observed sSFRs for fixed stellar masses at $z\approx 0$ (see for instance \citealt{Mitchell2014}). The effect is seen clearly in the two largest halo mass bins and at stellar masses above log \Mstarnospace/\Msol \space \squiggle 10.5. We expect that the low gas fractions are due to hot gas stripping being too severe, leading to starvation and quenching that is too strong. Note that GP14+GRP is too gas poor even without the inclusion of ram-pressure stripping of the cold gas which, as stated previously, has been shown to be an important driver of gas removal in this regime (i.e. cluster scales). At high stellar masses, depletion of gas is caused by AGN feedback being too strong, not allowing further cooling and replenishment of the ISM.

When we study \HI gas fraction as a function of sSFR we find GP14+GRP delivers a better overall agreement with our measurements than gas fraction as a function of stellar mass. However, the predicted population of galaxies with very low sSFRs - and likely low gas fractions - are not visible in the parameter space shown. The predictions for haloes of log M$_{\rm h}/$\Msol \space $<$ 13.5 (blue and green lines) are particularly successful when we compare with our observations. At higher halo masses, the model predicts \HI gas fractions at fixed sSFR that are slightly too high compared to our observations. This is likely due to GP14+GRP not accounting for the ram-pressure stripping of \HInospace, which is expected to significantly drive down gas fractions in this regime at both fixed sSFR and stellar mass. However, the inclusion of this effect would potentially increase tension between GP14+GRP and our observations.

For the comparison between observations and GP14+GRP, we tested the effect of using halo masses assigned via the abundance matching method rather than using the halo masses from the simulation on the scaling relations presented here. We follow the method of \citet{Yang2007}, calculating the total stellar mass from all the galaxies in a halo that have absolute {\it r}-band magnitude M$_{r}$ - 5 log (h) $\leq$ -19.5, ranking the groups using their integrated stellar mass and assigning a halo mass under the assumption that there is a one-to-one correspondence between the integrated stellar mass and halo mass. We found that the abundance matching method slightly under-predicts (\squiggle 0.1 dex) gas fractions at fixed stellar mass for halo masses below log M$_{\rm h}$/\Msol \space = 13.5, increasing the disagreement with observations. However, gas fractions at fixed sSFR show no significant differences between the samples using the two different halo masses.

Figures \ref{fig:models}c and \ref{fig:models}f show a comparison between the \HI fraction of satellites, as a function of halo mass, between the simulations of \citet{Dave2013} and observations, against stellar mass (\ref{fig:models}c) and sSFR (\ref{fig:models}f). The strongly bimodal distribution of satellite sSFRs in the \citet{Dave2013} model mean that quenched galaxies lie off the parameter space of Figure \ref{fig:models}f to lower sSFRs. We compute the average gas fraction of these galaxies (coloured arrows) and set them to our observational sSFR limit of log sSFR/yr = -13. \HI fractions of satellites in the simulation are systematically low (\squiggle 0.6 dex) when compared to observations and, while some qualitative agreement exists, there is limited reproduction of the general trends with stellar mass and sSFR as a function of halo mass.

The fact that disagreement is present even in low mass halos where stripping is inefficient in these simulations \citep{Rafieferantsoa2015} suggests that the origin of this deficit is likely endemic to the satellite population in the \citet{Dave2013} simulations. However, the explanation for this deficit is not straightforward, thus, below we briefly outline various possible causes, both physical and numerical.

Although the \HI content is greatly underestimated, the depletion of gas as a function of halo mass in the simulation somewhat echoes the observed trend, suggesting that key \HI removal processes are roughly followed. As shown in \citet{Rafieferantsoa2015}, at fixed halo mass, satellite \HI masses deviate from their stellar mass-matched central already at log M$_{\rm h}/$\Msol \space \squiggle11.5, while the data seems to suggest that such deviations do not begin until higher halo masses. This discrepancy may owe to overly-aggressive stripping or starvation within fairly low-mass halos, which then propagates to higher masses. Further to this, previous work using high resolution, non-cosmological simulations has shown stripping scenarios to be strongly dependent on both galaxy structure and the ISM model employed \citep{Mastropietro2005,Marcolini2003}.

From a numerical perspective, replicating the hydrodynamical interaction between \HI and the surrounding intracluster or intragroup medium is a notoriously difficult task. While remaining a dramatical improvement on the detail afforded by semi-analytics, the moderate resolution of \citet{Dave2013} simulations mean there is a possibility of dark matter particles being spuriously heated due to two-body interaction, this leads to artificial heating and momentum transfer to gas particles in the simulation \cite[see][]{Steinmetz1997,Abadi1999}. In addition and as briefly mentioned in Section \ref{sec:ezw_desc}, the version GADGET used does not include newer SPH recipes that are required to correctly capture the fluid instabilities at this interface. While this is an undoubted shortcoming of the simulation, it results in the employed SPH underestimating the effect of ram-pressure as well as other suppression mechanisms (i.e. viscous stripping, Kelvin-Helmholtz instabilities), thus discrepancies are expected to worsen with the inclusion of new SPH, not improve. \citet{Schaller2015} compared the old SPH formulation with newer SPH in the EAGLE \cite[Evolution and Assembly of GaLaxies and their Environments; ][]{Schaye2015,Crain2015} simulations and found that the amount of cold gas and therefore the star formation rates of galaxies are reduced in the new SPH formulation, evidencing that cold gas fractions would become even lower than those found in our comparison. However, \citet{Schaller2015} also show that other numerical aspects, such as the timestep limiter and how the sub-grid physics modules are implemented have significant effects on the properties of galaxies. The latter means that it is not straightforward to estimate how much the cold gas fractions may be affected in the \citet{Schaller2015} simulations, and instead direct testing is necessary in the future. 

Having compared our observations to theory, we see that models and simulations are producing far too many gas poor galaxies. The results show that considerable modifications are required if we are to successfully characterise the impact of environment on the \HI content of galaxies.

\section{Discussion and Conclusions}
\label{sec:Discussion}
In this paper we apply the \HI spectral stacking technique to study the effect of environment on the gas content of 10,567 satellite galaxies, selected by redshift and stellar mass from the intersection of SDSS and ALFALFA. We quantify environment using FoF, nearest neighbour and fixed aperture metrics. The FoF-based DM halo masses and group association are provided by the Y07 galaxy group catalogue, while 7th nearest neighbour and fixed aperture densities are computed separately. The main conclusions of this work can be summarised as follows:

\begin{itemize}
	\item Satellite galaxies in more massive haloes have, on average, lower \HInospace-to-stellar mass ratios at fixed stellar mass, surface density and sSFR than those in smaller haloes. The significant and systematic decrease in the gas content of satellites as a function of halo mass occurs across the entire group regime as well as the cluster environment. 

	\item Following this, we suggest that the average timescale for \HI loss from satellites in haloes with masses above log M$_{\rm h}$/\Msol \space $\geq$ 13 is considerably faster than the subsequent quenching of star formation.
	
	\item Gas content is also depleted with increasing nearest neighbour and fixed aperture densities. However, halo mass is the dominant environmental driver of \HI removal in satellites.
	
	\item Comparing our results to the predictions of theoretical models we find that, at fixed stellar mass and sSFR, both semi-analytics and hydrodynamic produce too many gas poor satellites. There is, however, qualitative agreement that the gas content of satellite galaxies depends upon the mass of their DM halo and depletion, particularly at fixed sSFR, is caused by a stripping mechanism.
\end{itemize}

We show that significant and continuous suppression of satellite \HI content due to environment is present in haloes more massive than log M$_{\rm h}$/\Msol\space \squiggle12 at fixed mass and morphology, and above log M$_{\rm h}$/\Msol\space \squiggle13 at fixed sSFR. By controlling for influences of mass, morphology and star formation, and separating those from the effect of environment, we present a scenario whereby environment driven processes are directly acting upon the \HI reservoirs of satellites. Observations have previously shown mechanisms such as strangulation, interaction and stripping to be prevalent in large galaxy groups and clusters. In order to explain our result of decreasing gas fractions in haloes of log M$_{\rm h}$/\Msol\space $\geq$ 12 we suggest that one or more of these processes becomes efficient in small to mid-size groups, well before the large group and cluster regimes. Encouragingly, this is entirely consistent with the results of \citet{Stark2016} who show a decrease in satellite gas content in halos above log M$_{\rm h}$/\Msol\space \squiggle12 at fixed stellar mass using individual detections.

From the evidence presented it is possible to take this analysis a step further and speculate upon the prominence of these different mechanisms, a subject that is still very much up for debate. To do so we divide the processes into two categories: slow acting, such as strangulation or starvation, on the one hand; and fast acting, which primarily refers to ram-pressure stripping, on the other. The path that galaxies trace through the gas fraction-sSFR plane (Figure \ref{fig:main_relations}b) when under the influence of these two categories differs considerably. A slow reduction in atomic gas naturally eventuates in a reduction of the molecular phase as H$_{2}$ is consumed by star formation and not replenished. This results in a steady decline in star formation \citep[$\gtrsim$ 1 Gyr, ][]{Balogh2000} as the available fuel reservoirs diminish and galaxies transition simultaneously to low gas fractions and onto the red sequence. In Figure \ref{fig:main_relations}b, this is seen in haloes with log M$_{\rm h}$/\Msol \space $ <$ 13 where differences in gas content as a function of {\it both} halo mass and sSFR are not significant.

In contrast, one would expect that when removal of \HI is (nearly) instantaneous \citep[$\gtrsim$ several 10 Myr, ][]{Vollmer2012}, there is a necessary time lag before star formation is quenched accordingly and the chosen star formation indicator registers a change. Galaxies that undergo such an event as they move into larger haloes will exhibit lower gas fractions at fixed sSFR between halo mass bins. It is for this reason that ram-pressure stripping is our preferred explanation for the lower gas fractions seen in haloes above log M$_{\rm h}$/\Msol \space $=$ 13 at fixed sSFR.

We caution the reader that, as stated in Section \ref{sec:Sample}, this paper uses two SFR estimators: fits to the H$\alpha$ emission lines from star forming galaxies and, where emission in H$\alpha$ is low, the break in galaxy spectra around 4000 \AA. While the H$\alpha$ emission traces star formation over the last few tens of Myr, the D4000 measurement is sensitive on longer timescales of \squiggle 1 Gyr. This means that, where the D4000 is used, we may simply conclude that the gas depletion occurs at a faster rate than 1 Gyr at fixed sSFR. However, for star forming galaxies, depletion of \HI with environment at fixed sSFR is indicative of an even more rapid removal of gas and ram-pressure stripping is favoured. This effect contributes to the broadening of the gas fraction-sSFR relation seen between the blue and red sequence in Figure \ref{fig:main_relations}b. 

In addition to FoF based halo masses, we also conducted our analysis using 7th nearest neighbour and fixed aperture densities. We have shown that, for the satellite population within our sample, there is a tight correlation between halo mass and local density metrics and, in Figure \ref{fig:fa_sat}, confirm that galaxies in denser environments tend to be more gas poor. We test whether gas is preferentially depleted by processes pertaining to either the halo mass (hydrodynamical) or the proximity of neighbours (gravitational) in Figure \ref{fig:mh_n7_sat}, showing that, at fixed stellar mass and sSFR, the reduction in \HI follows an increase in halo mass and over local density. This paints a physical picture where halo mass is the dominant factor in environment driven cold gas depletion and agrees with our hypothesis above - ram-pressure stripping is the likely candidate for \HI removal in massive haloes (log M$_{\rm h}$/\Msol \space $\geq$ 13). 

Having said this, it is important to check for the possibility of mergers driving the observed gas depletion with halo mass. In particular, for systems where the group-to-satellite mass ratio is low, dynamical friction timescales are short and mergers can occur rapidly. We discuss the impact of this effect on our results in Appendix \ref{Ap:A}.

The theoretical comparison presented in Figure \ref{fig:models} goes some way towards supporting the picture presented by observations. However, this work highlights the tendency for cosmological models to produce satellite populations that are too gas poor and therefore excessively quenched.

This comparison clearly illustrates that, while general trends of gas content with stellar mass and sSFR are grossly reproduced, significant improvements to both semi-analytic (GP14+GRP) and hydrodynamical (`ezw') approaches are required to fully match observations.

Recently, \citet{Marasco2016} conducted an analysis of environmental processes that affect the \HI content of galaxies in the EAGLE simulations that explicitly addresses some of the numerical concerns raised here. In their work, \citet{Marasco2016} find the fraction of \HI poor galaxies increases with halo mass, predicting an environmentally-driven bimodal distribution in the \HInospace-to-stellar mass ratio. The authors find that the most common environmental driver of gas in EAGLE satellites at z = 0 is ram-pressure stripping, with tidal forces and satellite-satellite interaction playing a secondary, yet significant role.

To conclude, this paper has provided evidence that the gas content of satellites is depleted by external processes as they transition into higher mass haloes and denser environments. Using observations and theory, the paper discusses the likely processes at work, suggesting that fast acting hydrodynamical mechanisms such as ram-pressure stripping are efficient in the group environment as well as the high density clusters. The results also exemplify the gains to be made by using the \HI spectral stacking technique, especially as we look to pave the way for the next generation of radio telescopes that will address these important questions.

\section*{Acknowledgements}
The authors wish to thank the referee for their helpful comments and constructive suggestions. We are grateful to K. Gereb, S. Janowiecki and A.R.H. Stevens for insightful discussions when writing this paper. We also acknowledge the work of the entire ALFALFA team in observing, flagging, and processing the ALFALFA data that this work makes use of. The ALFALFA team at Cornell is supported by NSF grants AST-0607007 and AST-1107390 and by the Brinson Foundation.

BC is the recipient of an Australian Research Council Future Fellowship (FT120100660). BC and LC acknowledge support from the Australian Research Council's Discovery Projects funding scheme (DP150101734).

Funding for SDSS-III has been provided by the Alfred P. Sloan Foundation, the Participating Institutions, the National Science Foundation, and the U.S. Department of Energy Office of Science. The SDSS-III web site is \url{http://www.sdss3.org/}.

SDSS-III is managed by the Astrophysical Research Consortium for the Participating Institutions of the SDSS-III Collaboration including the University of Arizona, the Brazilian Participation Group, Brookhaven National Laboratory, Carnegie Mellon University, University of Florida, the French Participation Group, the German Participation Group, Harvard University, the Instituto de Astrofisica de Canarias, the Michigan State/Notre Dame/JINA Participation Group, Johns Hopkins University, Lawrence Berkeley National Laboratory, Max Planck Institute for Astrophysics, Max Planck Institute for Extraterrestrial Physics, New Mexico State University, New York University, Ohio State University, Pennsylvania State University, University of Portsmouth, Princeton University, the Spanish Participation Group, University of Tokyo, University of Utah, Vanderbilt University, University of Virginia, University of Washington, and Yale University.

\bibliographystyle{mn2e}
\bibliography{refs}

\appendix
\section{The Effect of Mergers on Gas Fraction}
\label{Ap:A}

\begin{figure*}
\begin{multicols}{2}
    \includegraphics[width=\linewidth]{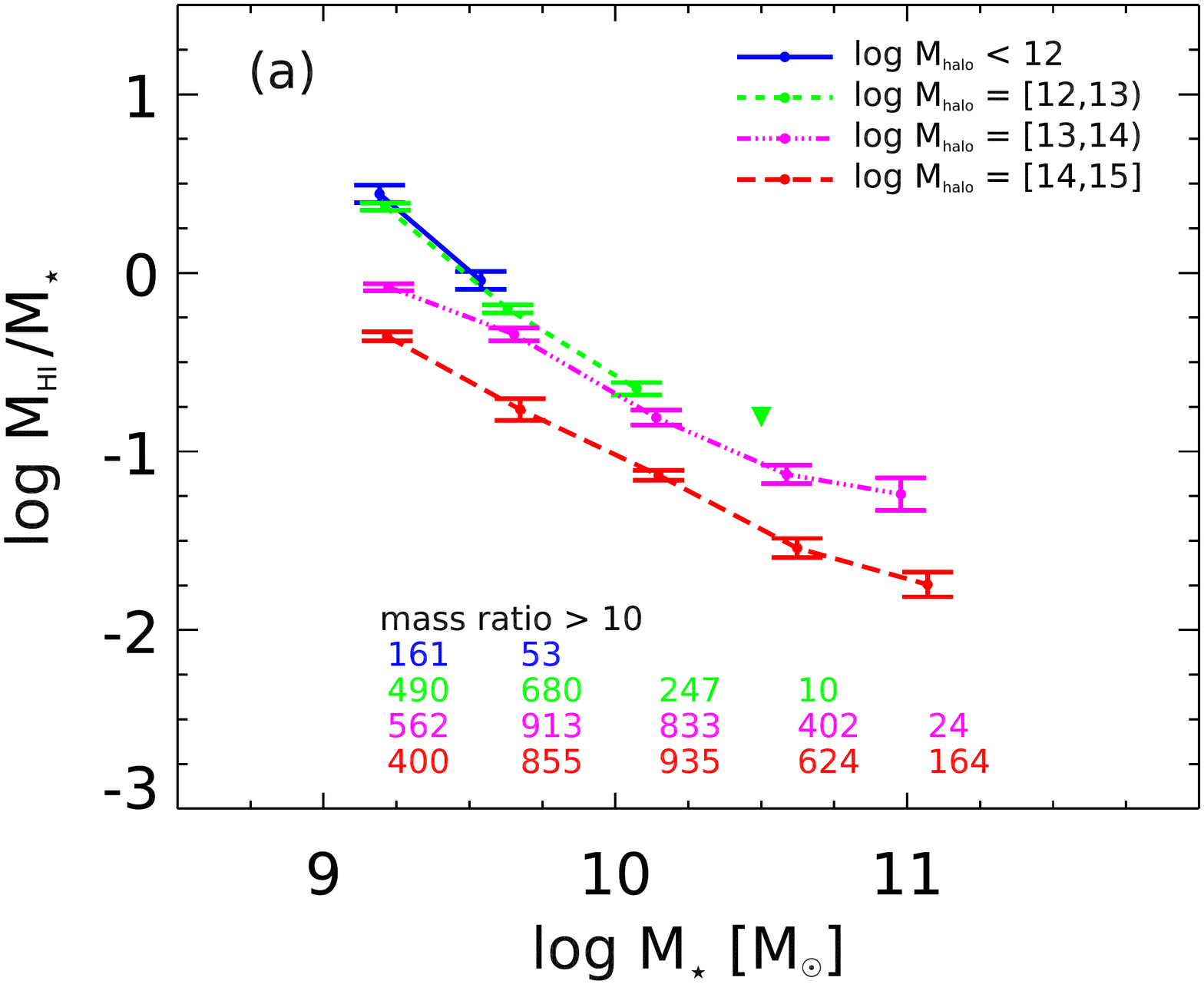}\par 
    \includegraphics[width=\linewidth]{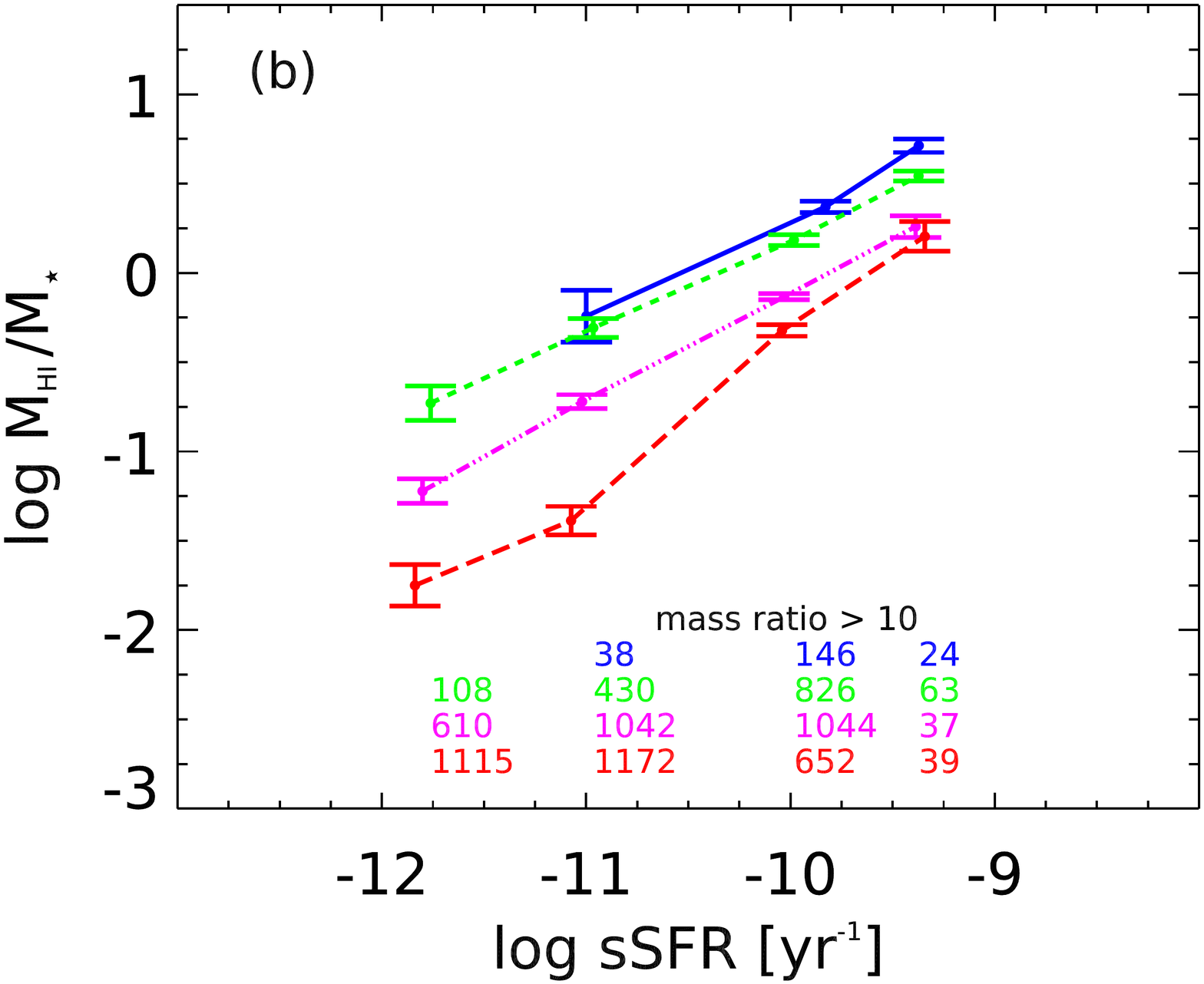}\par 
    \end{multicols}
  \caption[LoF entry]{Log \MHInospace/M$_{\star}$ versus log M$_{\star}$ (a) and log sSFR (b) for Sample I satellite galaxies where M$_{\star,grp}$/M$_{\star,sat} > 10$. This subset is binned according to the halo mass limits given in the legend. Stacked non-detections are denoted by triangles.}
\label{fig:mass_ratio}
\end{figure*}

Satellites with a mass similar to that of their group are subject to short dynamical friction timescales and thus increased merger rates \citep[see ][]{Chandrasekhar1943,Weinberg1998,Colpi1999,Taffoni2003}. In order to verify our definition of a `satellite' and discriminate between the stripping scenario (as outlined above) and galaxy mergers in driving gas content, we check our results excluding satellites with a ratio between group total stellar mass and satellite mass (M$_{\star,grp}$/M$_{\star,sat}$) less than ten. This is shown in Figure \ref{fig:mass_ratio} where we plot the gas fraction-stellar mass and sSFR scaling relations, binned by halo mass, for this subset of Sample I (7353 satellites). Despite approximately 30 per cent of satellites residing below this limit, the trend of gas depletion at fixed stellar mass and sSFR remains once these galaxies are removed from the sample. We also note that the rate of confusion within our sample is less than 10 per cent and satellites with a short dynamical friction timescale would most likely be flagged as confused. We therefore rule out mergers as the main driver of gas depletion due to environment at fixed stellar mass and sSFR.
\end{document}